\documentclass[structabstract]{aa}
\usepackage{savesym}
\usepackage{amsmath}
\savesymbol{iint}
\usepackage{txfonts}
\restoresymbol{TXF}{iint}
\usepackage{graphics,graphicx}
\usepackage{amssymb}
\usepackage{color,hyperref}
\usepackage{breakurl}
\usepackage{array}
\usepackage{rotating}
\usepackage{xcolor}
\usepackage{natbib}
\usepackage{float}
\usepackage{flafter}
\usepackage{booktabs}
\usepackage{afterpage}
\usepackage{lipsum}
\usepackage{longtable,pdflscape}
\bibpunct{(}{)}{;}{a}{}{,}

\newcommand{\SB}{{30~Dor~C}}
\newcommand{\SNR}{{MCSNR~J0536$-$6913}}
\newcommand{\chandra}{{\it Chandra}}
\newcommand{\xmm}{{\it XMM-Newton}}

\newcommand{\suzaku}{{\it Suzaku}}

\newcommand{\vpshock}{{\tt vpshock}}
\newcommand{\vphabs}{{\tt vphabs}}
\newcommand{\vapec}{{\tt vapec}}
\newcommand{\phabs}{{\tt phabs}}
\newcommand{\pow}{{\tt powerlaw}}
\newcommand{\srcut}{{\tt srcut}}

\begin{document}

\title{{\it XMM-Newton} study of \SB\ and a newly identified \SNR\ in the Large Magellanic Cloud\thanks{Based on observations obtained with \xmm, an ESA science mission with instruments and contributions directly funded by ESA Member States and NASA}}



\author{P.~J.~Kavanagh \inst{1} \and M.~Sasaki \inst{1} \and L.~M.~Bozzetto \inst{2} \and M.~D.~Filipovi\'c \inst{2} \and S.~D.~Points \inst{3} \and P.~Maggi \inst{4} \and F.~Haberl \inst{4}}

 

\institute{Institut f\"{u}r Astronomie und Astrophysik, Kepler Center for Astro and Particle Physics, Eberhard Karls Universit\"{a}t T\"{u}bingen, Sand 1, T\"{u}bingen D-72076, Germany\\ \email{kavanagh@astro.uni-tuebingen.de}
\and University of Western Sydney, Locked Bag 1797, Penrith South DC, NSW 1797, Australia
\and Cerro Tololo Inter-American Observatory, Casilla 603, La Serena, Chile
\and Max-Planck-Institut f\"{u}r extraterrestrische Physik, Giessenbachstra\ss e, D-85748 Garching, Germany}

\date{Received ?? / Accepted ??}

\abstract{}{We present a detailed study of the superbubble \SB\ and the newly identified \SNR\ in the Large Magellanic Cloud.}{All available \xmm\ data (flare-filtered exposure times of 420~ks EPIC-pn, 556~ks EPIC-MOS1, 614~ks EPIC-MOS2) were used to characterise the thermal X-ray emission in the region. An analysis of the non-thermal X-ray emission is also presented and discussed in the context of emission mechanisms previously suggested in the literature. These data are supplemented by X-ray data from \chandra, optical data from the Magellanic Cloud Emission Line Survey, and radio data from the Australia Telescope Compact Array and the Molonglo Observatory Synthesis Telescope.}{The brightest thermal emission towards \SB\ was found to be associated with a new supernova remnant, \SNR. X-ray spectral analysis of \SNR\ suggested an ejecta-dominated remnant with lines of O, Ne, Mg, and Si, and a total $0.3-10$~keV X-ray luminosity of $\sim8 \times 10^{34}$ erg~s$^{-1}$. Based on derived ejecta abundance ratios, we determined the mass of the stellar progenitor to be either $\sim18$~M$_{\sun}$ or as high as $\gtrsim40$~M$_{\sun}$, though the spectral fits were subject to simplifying assumptions (e.g., uniform temperature and well-mixed ejecta). The thermal emission from the superbubble exhibited enrichment by $\alpha$-process elements, evidence for a recent core-collapse SNR interaction with the superbubble shell. We detected non-thermal X-ray emission throughout \SB, with the brightest regions being highly correlated with the H$\alpha$ and radio shells. We created a non-thermal spectral energy distribution for the north-eastern shell of \SB\ which was best-fit with an exponentially cut-off synchrotron model.}{Thermal X-ray emission from \SB\ is very complex, consisting of a large scale superbubble emission at the eastern shell wall with the brightest emission due to \SNR. The fact that the non-thermal spectral energy distribution of the superbubble shell was observed to roll-off is further evidence that the non-thermal X-ray emission from \SB\ is synchrotron in origin.}
\titlerunning{{\it XMM-Newton} study of \SB}
\maketitle 

\section{Introduction}
Massive stars, via stellar winds and later supernovae (SN), are responsible for energising and enriching the interstellar medium (ISM). The collective mechanical output of massive star clusters into the surrounding ISM creates so-called ÔsuperbubblesÕ (SBs), 100-1000 pc diameter shells of swept-up interstellar material which contains a hot ($10^{6}$ K), shock-heated gas \citep{Weaver1977,MacLow1988}. In recent years, non-thermal X-ray emission has been detected from a number of Galactic [\object{RCW~38} \citep{Wolk2002}, \object{Westerlund~1} \citep{Muno2006}] and extragalactic SBs [\object{N11} \citep{Maddox2009}, \object{N51D} \citep{Cooper2004}, \object{30~Dor~C} \citep{Bamba2004,Smith2004,Yam2009} in the \object{Large Magellanic Cloud} (LMC); and \object{IC131} in \object{M33} \citep{Tullmann2009}]\footnote{We note that the non-thermal detection in LMC SBs \object{N11}, \object{N51D}, and \object{N70} has been called into question \citep{Yam2010,DeHorta2014}.}.

\par Proposed explanations for the non-thermal emission detected in SBs are synchrotron radiation, inverse Compton (IC) scattering of ambient photons, and non-thermal Bremsstrahlung radiation. A prerequisite for any of these mechanisms is the acceleration of electrons to relativistic or near relativistic energies. \citet{Parizot2004} demonstrated that turbulence and magnetohydrodynamic waves born out of colliding massive stellar winds and SN shocks can accelerate and re-accelerate particles to such energies. \citet{Butt2008} also argue that the energy loss to relativistic particle acceleration can be invoked to explain the SB energy discrepancy, the persistent problem that the observed combined thermal and mechanical energy in SBs is less than the total energy input of the stellar population \citep[see][for examples]{Cooper2004,Maddox2009,Kavanagh2012}. \SB\ is by far the strongest non-thermal X-ray emitting SB in the LMC, and thus provides an ideal laboratory for probing the non-thermal emission mechanisms of, and resulting effects on SBs and their evolution.



\noindent \SB\ is located to the southwest of the main 30 Dor complex and was first identified (and named) by \citet{LeMarne1968} and later classified as an SB by \citet{Mathewson1985} using radio and optical emission line data. The SB is powered by the \object{LH~90} \citep{Lucke1970} OB association which consists of several clusters \citep[ages from $\sim3-7$ Myr,][]{Testor1993}. Discussion of the radio and H$\alpha$ shells of \SB\ can be found in \citet{Mathewson1985} and \citet[][henceforth SW04]{Smith2004}. The first X-ray detection of \SB\ was with {\it Einstein} \citep{Long1981}. The SB later had a place in the history of X-ray astronomy, being observed in the first-light \xmm\ observation \citep{Dennerl2001}, presenting a ring-like structure in hard X-rays unlike all other extended sources in the \object{LMC}. 

\par Observations with the current generation of X-ray missions (\chandra, \xmm\ and \suzaku) have provided a wealth of information on this object. \citet[][henceforth BU04]{Bamba2004} reported on the analysis of two \chandra\ ACIS-S and two early \xmm\ observations of \SB. Power-law fits to the non-thermal shell emission resulted in best-fit photon indices ($\Gamma$) in the range of $2.1-2.9$, indicative of a synchrotron origin, and the authors conclude that this is the emission mechanism. Contemporaneously, SW04 reported an analysis of the same \xmm\ observations as BU04. However, SW04 argue that the synchrotron mechanism cannot be the origin of the non-thermal X-rays based on energetics considerations. The expansion of the bubble is much too slow to produce the high energy particles required for non-thermal X-ray synchrotron emission in the shell. SW04 also considered IC and non-thermal Bremsstrahlung mechanisms as the source of the hard X-rays. While IC scattering of cosmic microwave background (CMB) and IR photons by relativistic electrons from a young pulsar wind could explain the non-thermal X-rays, there is, as yet, no solid observational evidence for such a source of high energy particles in \SB. Non-thermal Bremsstrahlung was found to be too inefficient a process. As with BU04, \citet[][henceforth YB09]{Yam2009} suggested a synchrotron origin to the non-thermal X-rays, based on the statistical rejection of a simple power law over an \srcut\footnote{Synchrotron spectrum from an exponentially cut off power-law distribution of electrons in a homogeneous magnetic field \citep{Reynolds1998}.}  model. They posited that an SNR from deep in the bubble has freely expanded through the interior and is now interacting with the SB shell walls.  However, this interpretation presents problems with regard to standard SB theory. First of all, and as discussed in SW04, there are no shocks fast enough at the shell to sustain this mechanism. Secondly, a remnant does not freely expand from deep in an SB interior to the shell wall since its energy is dissipated by turbulence long before this \citep{MacLow1988,Parizot2004}. 

\par In addition to the non-thermal shell, all previous authors found that the southeastern regions of \SB\ exhibited substantial thermal emission. Thermal plasma models with enhanced $\alpha$-process elements were required to adequately fit this emission. Due to these metal enhancements, it has been suggested that the emission is the result of a recent SNR interaction with the shell wall. However, there has yet to be a detailed spatially resolved spectral analysis of the thermally emitting regions in \SB.

\par In this paper we seek to carry out a comprehensive study of the non-thermal and thermal X-ray emission in \SB\ using the ample archival \xmm\ data. Several hundred ks of \xmm\ data has been collected in recent years due to \SB\ being located only a few arcmins from \object{SN~1987A}, which has been the subject of a deep monitoring campaign \citep{Heng2008,Sturm2010,Maggi2012}. In addition to this abundance of \xmm\ data, we have new radio observations of \SB\ with the Australia Telescope Compact Array (ATCA), supplementing already available radio data from the Molonglo Observatory Synthesis Telescope (MOST), and optical emission line data from the Magellanic Cloud Emission Line Survey \citep[MCELS][]{Smith2006}. Using these multi-wavelength data our goal is to obtain a clear picture of the physical processes and mechanisms at work in this intriguing object. In Section~\ref{odr} we outline the multi-wavelength observations and data reduction. In Section~\ref{an} we describe the detailed analysis of the observational datasets. In Section~\ref{dis} we discuss the results of our analysis of the thermal and non-thermal X-ray emission in \SB\ in the context of the multi-wavelength picture before giving a summary of our work in Section~\ref{sum}.

\section{Observations and data reduction}
\label{odr}
\subsection{Optical}
We used images obtained during the MCELS \citep{Smith2006}, taken with the 0.6 m University of Michigan/Cerro Tololo Inter-American Observatory (CTIO) Curtis Schmidt Telescope which is equipped with a SITE 2048 $\times$ 2048 CCD, producing individual images of $1.35^{\circ} \times 1.35^{\circ}$ at a scale of 2.3$\arcsec$ pixel$^{-1}$. The survey mapped both the LMC ($8^{\circ} \times 8^{\circ}$) and the Small Magellanic Cloud ($3.5^{\circ} \times 4.5^{\circ}$) in narrow bands covering [\ion{O}{iii}]$\lambda$5007 \AA, H$\alpha$, and [\ion{S}{ii}]$\lambda$6716, 6731 \AA, in addition to matched green and red continuum bands. The survey data were flux calibrated and combined to produce mosaicked images. We extracted cutouts centred on \SB\ from the MCELS mosaics. We subtracted the continuum images from the corresponding emission line images, thereby removing the stellar continuum and revealing the full extent of the faint diffuse emission. We note here that SW04 used the MCELS H$\alpha$ data to aid in their discussion of the morphological properties of \SB.

\subsection{Radio}
\label{radio-obs}
Radio-continumm data used in this project includes a 36 cm (843 MHz) MOST mosaic image \citep[as described in][]{Mills1984} and complementary 20~cm (1380 MHz) observations from the ATCA project C221 (PI: J.~M.~Dickey). These observations include three pointings in the vicinity of \SB, which were mosaicked to gain a higher quality image of the region. Details of these observations are listed in Table \ref{rc-obs}. We used the \textsc{miriad}\footnote{\burl{http://www.atnf.csiro.au/computing/software/miriad/}}  \citep{Sault1995} and \textsc{karma} \citep{Gooch1995} software packages for reduction and analysis. We created images using \textsc{miriad} multi-frequency synthesis \citep{Sault1994} and natural weighting. They were deconvolved with primary beam correction applied. The same procedure was used for both \textit{U} and \textit{Q} stokes parameter maps. More information about the data reduction and a number of other LMC SNR studies can be found in \citet{Boj2007}, and reference therein.

\begin{table}
\caption{Radio-continuum observations used from ATCA project C221}
\begin{center}
\begin{tabular}{lcccc}
\hline
Date & Time & RA & Dec & Array \\
 & (min.) & & & \\
 \hline
1993-01-15/16	&	437.0	&	  5:38:47	&	-69:05:50	&	750B	\\
1993-01-15/16	&	427.7	&	  5:38:47	&	-69:27:50	&	750B	\\
1993-01-15/16	&	589.3	&	  5:35:27	&	-69:16:32	&	750B	\\
1993-01-28/29	&	410.7	&	  5:38:47	&	-69:05:50	&	750A	\\
1993-01-28/29	&	403.4	&	  5:38:47	&	-69:27:50	&	750A	\\
1993-01-28/29	&	572.0	&	  5:35:28	&	-69:16:32	&	750A	\\
1993-03-13/14	&	643.4	&	  5:38:47	&	-69:05:50	&	1.5D		\\
1993-03-13/14	&	639.7	&	  5:38:47	&	-69:27:50	&	1.5D		\\
1993-03-13/14	&	868.4	&	  5:35:28	&	-69:16:32	&	1.5D		\\
1993-05-08	&	642.3	&	ÊÊ5:38:47	&	-69:05:50	&	1.5A		\\
1993-05-08	&	634.3	&	ÊÊ5:38:47	&	-69:27:50	&	1.5A		\\
1993-05-08	&	854.0	&	ÊÊ5:35:28	&	-69:16:32	&	1.5A		\\
\hline
\end{tabular}
\end{center}
* Observations were taken at a  frequency of 1380 MHz using a bandwidth of 128 MHz
\label{rc-obs}
\end{table}

\subsection{X-ray}
\subsubsection{\xmm}
We obtained all of the data on \SB\ available from the the \xmm\ Science Archive, consisting of 15 observations spread across 12 years. We assessed each of the observational datasets for their suitability to our analysis. We omitted observations which had flare-filtered exposure times (see Section \ref{x-ray-imaging}) less than 10~ks leaving 11 observations which we used for our analysis. These observations and the flare-filtered exposure times are listed in Table \ref{observations}.

\begin{table}
\caption{\xmm\ observations of \SB\ used in the analysis}
\begin{center}
\begin{small}
\label{observations}
\begin{tabular}{lllccc}
\hline
Obs. ID & Obs. Date & PI &	\multicolumn{3}{c}{Exposure time (ks)} \\
 &  & & pn & MOS1 & MOS2 \\
\hline
\hline
0104660101 & 2000-09-17 & Watson  & 22.3  & --  & --  \\

0104660301 & 2000-11-25 & Watson  & --  & 20.7  & 19.6  \\

0113020201 & 2001-11-19 & Aschenbach   & --  & 31.5  & 25.0  \\

0144530101 & 2003-05-10  & McCray & --  & 46.8  & 46.8  \\
	
0406840301 & 2007-01-17 & Haberl  & 53.3  &  74.4 &  76.0 \\

0506220101 & 2008-01-11 & Haberl  & 61.2  & 80.7  & 83.4  \\

0556350101 & 2009-01-30 & Haberl  &  57.3 & 79.0  & 81.7  \\

0601200101 & 2009-12-11 & Haberl  &  70.8 & 85.5  & 85.5  \\

0650420101 & 2010-12-12  & Haberl &  46.1 & 57.9  & 60.7  \\

0671080101 & 2011-12-02 & Haberl & 56.4  & 67.8  & 69.0  \\

0690510101 & 2012-12-11 & Haberl  & 52.9  & 11.9  & 66.7  \\

\hline
\multicolumn{5}{l}{All exposure times are flare-filtered exposure times. The}\\
\multicolumn{5}{l}{ target name for all observations was SN 1987A, except}\\
\multicolumn{5}{l}{ Obs. ID 0113020201, for which the target was}\\
\multicolumn{5}{l}{ \object{PSR~J0537$-$6909}.}\\

\end{tabular}
\end{small}
\end{center}
\end{table}%

\par The available data were collected over an extended period of time and were subject to varying instrumental performance and response. Hence, we required a consistent reduction and analysis method. In addition, the final science products should be free of as much background contaminants as possible to minimise the complexity of the analysis. Thus, we used the \xmm\ Extended Source Analysis Software (XMM-ESAS), packaged in SAS 12.0.1. XMM-ESAS is based on the software used for the background modelling described in \citet{Snowden2004}. Essentially, XMM-ESAS consists of a set of tasks to produce images and spectra from observational data, and to create model quiescent particle background (QPB) images and spectra which can be subtracted from the observational science products \citep[see][]{Kuntz2008,Snowden2008}. We processed each of the observational datasets according to the ESAS Cookbook\footnote{Available at \burl{http://heasarc.gsfc.nasa.gov/docs/xmm/xmmhp\_xmmesas.html}}. Standard filtering and calibration were applied using the SAS tools \texttt{epchain}, \texttt{emchain}, and the XMM-ESAS tools \texttt{pn-filter}, and \texttt{mos-filter}. The CCDs of each of the EPIC instruments were then examined to ensure that none were operating in an anomalous state \citep[where the background at $E < 1$ keV is strongly enhanced, see][]{Kuntz2008}. 

\par
\begin{figure*}
\begin{center}
\resizebox{\hsize}{!}{\includegraphics[trim= 0cm 0cm 0cm 0cm, clip=true, angle=0]{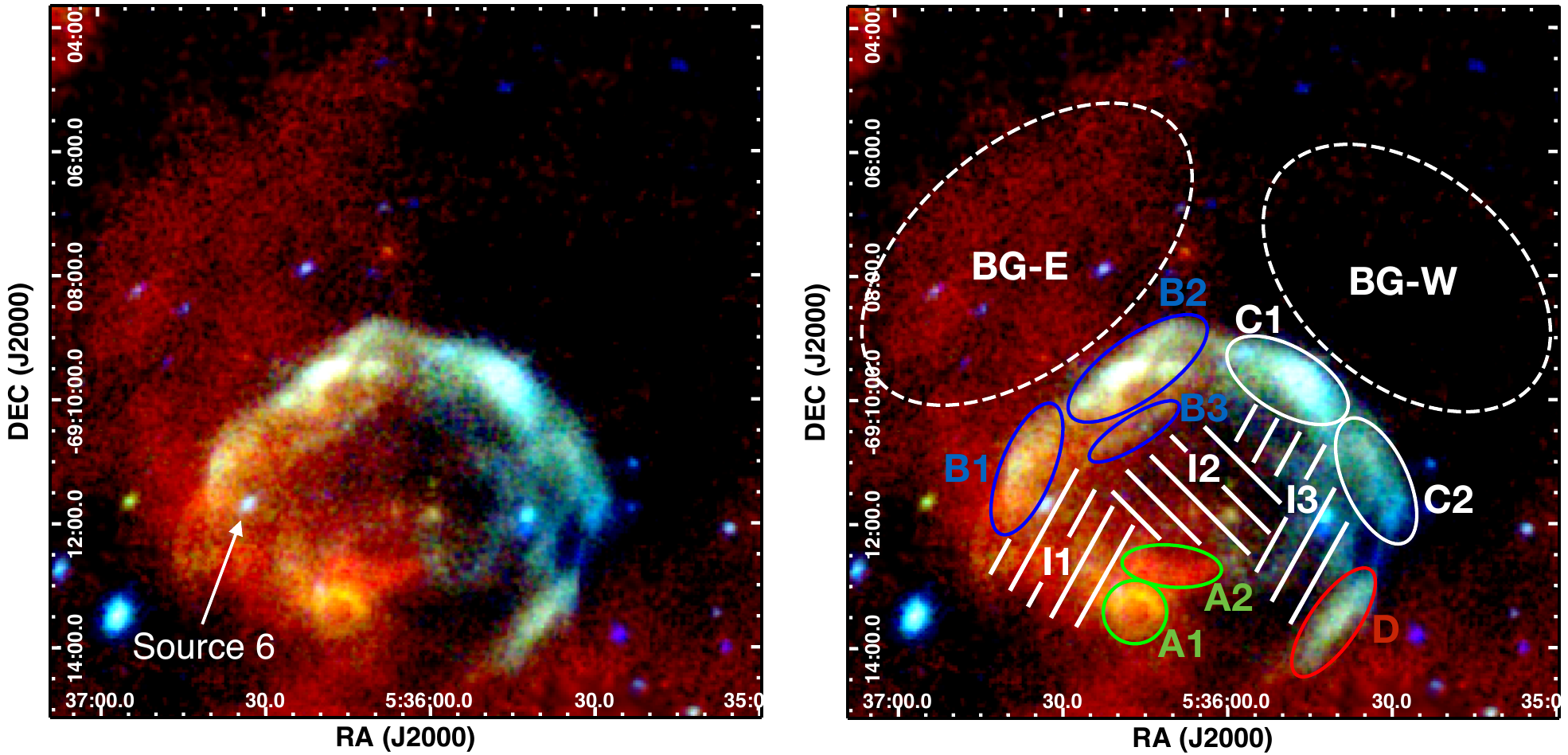}}
\caption{$Left$: Combined \xmm\ EPIC image of \SB\ in false colour with red, green, and blue corresponding to 0.3--1~keV, 1--2~keV, and 2--7~keV. Also shown is the region to the north of \SB\ which highlights the highly variable background in the region.  The image has been exposure corrected in each energy band and have the QPB subtracted, and binned into 2x2 pixel bins before being adaptively smoothed using the XMM-ESAS task \texttt{adapt-2000}. The position of Source 6, which is discussed as a possible compact object in Section~\ref{thermal-sb}, is also marked. $Right$: Same as $Left$ but with the spectral analysis regions indicated. See Section \ref{an-regions} for a description of the regions. 
}
\label{xmm-rgb}
\end{center}
\end{figure*}

 The \texttt{pn-spectra} and \texttt{mos-spectra} tasks were used to produce images in the $0.3-1$~keV, $1-2$~keV, and $2-7$~keV energy bands for each dataset. The \texttt{pn-back} and \texttt{mos-back} tasks were then used to produce corresponding QPB images. We then used \texttt{merge\_comp\_xmm} to create mosaicked count, exposure, and QPB images. Finally, the \texttt{adapt\_2000} task was implemented to create exposure corrected mosaics in each energy band with the QPB subtracted, bin them into 2x2 pixel bins, and adaptively smooth the resulting image. We combined these mosaics to produce an RGB image, shown in Fig. \ref{xmm-rgb}~($left$).

\par The XMM-ESAS task \texttt{cheese-bands} was used to search for point sources in the FOV. This task performs the source detection on all three EPIC instruments simultaneously in multiple energy bands. We selected the $0.3-1$~keV, $1-2$~keV, and $2-7$~keV bands so that the source detection would be sensitive to both soft and hard sources in the FOV. However, this task operated on individual observational datasets only and not on a final merged product. Thus, the source lists were examined using the final merged images to search for faint sources which were missed. Any sources found were added to the point source mask used for the spectral analysis using the SAS task \texttt{region} and the XMM-ESAS task \texttt{make\_mask}. 

\par For extracting spectra and response files, we used the XMM-ESAS tasks \texttt{pn-spectra} and \texttt{mos-spectra}. The tasks \texttt{pn-back} and \texttt{mos-back} were used to produce corresponding QPB spectra to be subtracted from the observational spectra. The spectra were grouped to minimum of 30 counts per bin to allow the use of the $\chi^{2}$-statistic. All fits were performed using XSPEC \citep{Arnaud1996} version 12.7.1 with ATOMDB\footnote{\burl{http://www.atomdb.org/}} version 2.0.1, abundance tables set to those of \citet{Wilms2000}, and photoelectric absorption cross-sections set to those of \citet{Bal1992}. For spectral components associated with the LMC, abundances were fixed to 50\% the solar value \citep{Russell1992}. We limit our analysis to the 0.4-7 keV range as recommended in the ESAS Cookbook. Thus, we avoid the strong low energy tail of the EPIC-pn quiescent background due to detector noise and the EPIC-pn fluorescence line forest just above 7 keV. In all the forthcoming fits, spectra which have been extracted from a given region from all the observational datasets are fit simultaneously in XSPEC. We consider only those EPIC spectra with $>1000$ QPB background subtracted counts to ensure the statistical quality of the spectra in the fits.

\subsubsection{\chandra}
\label{chan-red}
We also utilised the available \chandra\ \citep{Weiss1996} data to aid in the discussion in Section~\ref{new-snr}. The Advanced CCD Imaging Spectrometer S-array \citep[ACIS-S][]{Garmire2003} has observed \SB\ twice during observations of the nearby SN~1987A. These are ObsID~1044 ($\sim18$~ks, PI: G.~Garmire) and ObsID~1967 ($\sim99$~ks, PI: R.~McCray). For a detailed analysis of these observations with respect to \SB\ the reader is directed to BU04. We reduced and analysed the \chandra\ observations using the CIAO~v4.6.1 software package \citep{Fru2006} with CALDB~v4.5.9\footnote{Both available at \burl{http://cxc.harvard.edu/ciao/}}. Each data set was reduced using the contributed script \texttt{chandra\_repro}. Combined energy filtered and exposure corrected images were produced using the \texttt{merge\_obs} script.

\section{Analysis}
\label{an}
\subsection{X-ray morphology}
\label{x-ray-imaging}
\par The well-known non-thermal shell is seen in unprecedented detail (Fig.~\ref{xmm-rgb}), with structure visible in regions of stronger emission. In the S-SE region there is an obvious circular emission region, most notable in the $1-2$~keV energy range. The morphology and classification of this object is discussed in detail in Section~\ref{snr-class}. In addition, the X-ray background is not uniform, with a very obvious dichotomy between the east and west regions of \SB. The eastern side is projected against large scale hot ISM emission. This emission is much less apparent on the western side, most likely due to the known molecular clouds located in the foreground \citep[][BU04]{Johansson1998}. Due to the background variation, we must, as much as possible, take this into account when treating the background in the spectral analysis of \SB. 

\subsection{\SNR}
\label{snr-class}
\par An additional extended X-ray emitting object is evident as a circular shell in the $1-2$~keV band, projected against the southern \SB\ shell (see region~A1 in Fig.~\ref{xmm-rgb} {\it right}). Object classes that can produce diffuse X-ray structures in extragalactic observations are galaxy clusters, SBs, and SNRs \citep[see][for a more detailed description of the X-ray properties of these objects]{Maggi2014}. We ruled out the possibility that this object is hot gas in the intracluster medium of a background galaxy cluster since the observed shell morphology of the object is not in keeping with that expected from the hot gas of a galaxy cluster, which is centrally filled. It is also unlikely that this structure is an SB, since these require a high mass stellar population to drive their expansion, which is absent here. An SNR is a far more likely explanation given the shell morphology. Hence, we proceed with the assumption that the object is an SNR, and assess other tracers of this object classification.

\par Typically, objects are classified as SNRs based on satisfying certain observational criteria. For example, the Magellanic Cloud Supernova Remnant (MCSNR) Database\footnote{\burl{http://www.mcsnr.org/about.aspx}} state that at least two of the following three observational criteria must be met: significant H$\alpha$, [\ion{S}{ii}], and/or [\ion{O}{iii}] line emission with an [\ion{S}{ii}]/H$\alpha$ flux ratio $>0.4$ \citep{Mathewson1973,Fesen1985}; extended non-thermal radio emission; and extended thermal X-ray emission. A discussion on the significance of each of these classification criteria is given in \citet{Filipovic1998}.  The new candidate SNR satisfies only one of these three criteria, since \citet{Mathewson1985} found that [\ion{S}{ii}]/H$\alpha<0.4$ throughout \SB\ and our radio data show no clear indications of an SNR. These multi-wavelength properties are discussed in detail in Section~\ref{new-snr}. Even in the absence of optical and radio emission tracers, we are confident in classifying this object as an SNR given the 1--2~keV shell morphology and X-ray spectral signatures (see Sections~\ref{snr-fit} and \ref{new-snr}), and we hereafter refer to this source as \object{MCSNR~J0536$-$6913} (see forthcoming text for position determination).

\par The remnant's shell morphology is extraordinarily circular with a north-south brightness gradient. It is likely that the SNR is located outside of \SB\ rather than inside the SB. If the SNR was located inside, we would not expect to observe a shell morphology since the blast wave would only encounter a low density hot plasma. The brighter emission from the north of the SNR suggests it is evolving into a higher density medium than in the south, which again is counter-intuitive to a location in the bubble. If the SNR is located outside \SB\ but near enough that the northern shell is evolving towards the SB shell of higher density, the expected density gradient could explain the X-ray morphology. 

\par
\begin{figure}
\begin{center}
\resizebox{\hsize}{!}{\includegraphics[trim= 0.2cm 13.55cm 21.5cm 0.2cm, clip=true, angle=0]{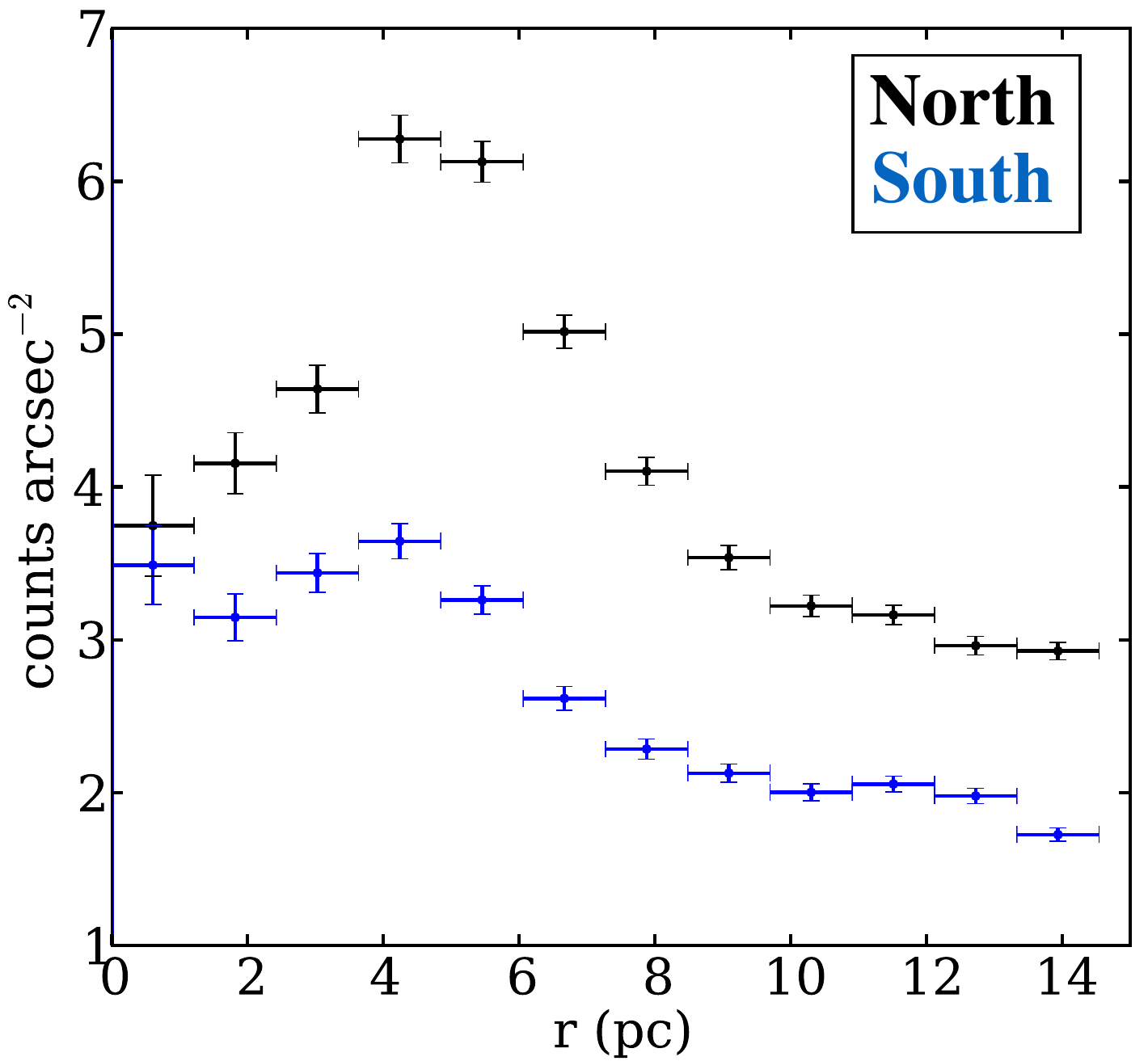}}
\caption{Northern (black) and southern (blue) radial profile of \SNR\ in the $1-2$~keV band.
}
\label{rad-pro}
\end{center}
\end{figure}

 As discussed later in Section \ref{snr-fit}, the notable shell in the $1-2$~keV band is likely shocked ejecta emission. While the outer edges of the ejecta are somewhat smeared out, the inner edge is much brighter and well defined. This may represent either the progression of the reverse shock into the ejecta or the radius at which the ejecta distribution has fallen to a level where ejecta emission is no longer detectable. We fit a circular region to the inner edge of the ejecta. We take the centre of this circle to be the position of the remnant, which gives a J2000 position of RA~=~05$^{\rm{h}}$36$^{\rm{m}}$17.0$^{\rm{s}}$ and Dec~=~$-69$$^{\rm{d}}$13$^{\rm{m}}$28$^{\rm{s}}$, leading to identifier \SNR. To estimate the extent of the SNR we follow a similar prescription to \citet{Maggi2014}. Firstly, we created radial profiles of the remnant. Because of the differing local backgrounds and brightness of the ejecta, we split the radial profiles into northern and southern components. The northern ejecta are much brighter than the southern and are immersed in a higher X-ray background due to the shell of \SB. The southern ejecta is evolving away from the SB emission and has a correspondingly lower background. We take radial bins of $5\arcsec$ out to an angular distance of $1\arcmin$, corresponding to $\sim1.2$~pc bins out to $\sim14.4$~pc at the LMC distance of 50~kpc \citep{diBen2008}, and determine the surface brightness of each bin. The north and south radial profiles are shown in Fig. \ref{rad-pro}. Following \citet{Maggi2014}, the dimensions of the remnant are defined where the intensity has fallen to 26\% of the peak following background subtraction. If the radial profile is Gaussian, this would enclose 90\% of the distribution. Taking the background from radial bins $>10$~pc, we determined the radii of the northern and southern shells to be $\sim8(\pm1)$~pc, by simply taking the first bin above the threshold. There are some caveats to be aware of with the determined SNR dimensions. First, with regard to the outer radius, the X-ray emission due to the ejecta may not represent the furthermost emission from the SNR centre. It is very likely that the forward shock has swept-up and shocked ISM, located ahead of the ejecta, but the spatial resolution of \xmm\ cannot resolve the two components. The outermost edge of the shell in the $1-2$~keV range more likely traces the contact discontinuity between ISM shocked by the blast wave and ejecta shocked by the reverse shock. Such an SNR structure is evident in high-spatial resolution images of similar MC remnants from \chandra~ \citep[e.g.][]{Warren2004,Sasaki2006}. 


\subsection{X-ray spectral analysis}
Following the subtraction of the modelled QPB spectra, the remaining particle induced background consists of instrumental fluorescence lines and the residual soft-proton (SP) contamination. The instrumental fluorescence lines can be modelled with Gaussian components (\texttt{gauss} in XSPEC) at 1.49 keV in the EPIC-pn spectrum, and 1.49 keV and 1.75 keV for the EPIC-MOS spectra. The level of residual SP contamination varies from observation to observation. We estimated the level of this using the diagnostic tool of \citet{DeLuca2004}\footnote{\burl{http://xmm2.esac.esa.int/external/xmm_sw_cal/background/epic\_scripts.shtml\#flare}}. This allows us to anticipate observations which may be strongly affected by residual SP contamination which can be accounted for in the spectral modelling. However, due to the significant straylight contamination in the southern FOV from the high-mass X-ray binary LMC~X-1, the results of the diagnostic tool are compromised. Instead, we ran test fits on spectra from each observational dataset individually to search for signs of SP contamination, which manifest as a significant deviation from continuum emission at higher energies. We found that only Obs.~ID~0113020201 was badly affected by residual SPs. As such, an additional power law component not convolved with the instrumental response was included in this observation only. This was achieved using the diagonal response files supplied in the XMM-ESAS CALDB.

\par Given the strong variation in the astrophysical X-ray background (AXB) of \SB\ noted in Section \ref{x-ray-imaging}, we decided to split the treatment of the AXB into east and west divisions, which we hereafter refer to as BG-E and BG-W, respectively. BG-E contains contributions from the AXB, as well as a bright  soft component, likely due to hot ISM in the LMC. This hot ISM is less apparent in BG-W. This could be due to the foreground molecular cloud which covers the western side of \SB, however, this assumption is simplistic as an intrinsic reduction in surface brightness is also possible. For our analysis we attempted to fit each of the background regions with a physically motivated model which could then themselves be included in the spectral fits to the \SB\ spectra. 

\par The AXB typically comprises four or fewer components \citep{Snowden2008,Kuntz2010}, namely the unabsorbed thermal emission from the Local Hot Bubble (LHB, $kT\sim$ 0.1 keV), absorbed cool ($kT\sim$ 0.1 keV) and hot ($kT\sim$ 0.25 keV) thermal emission from the Galactic halo, and an absorbed power law \citep[$\Gamma \sim 1.46$,][]{Chen1997} representing unresolved background active galactic nuclei (AGN). In cases of low Galactic foreground absorption, the LHB and absorbed cool Galactic halo emission are indiscernible and can be modelled as a single component. The normalisation of the background AGN component can be fixed to an equivalent of 10.5 photons~keV~cm$^{-2}$~s$^{-1}$~sr$^{-1}$, as recommended in the XMM-ESAS documentation. All thermal components were fit with the \texttt{apec} \citep{Smith2001} thermal plasma model in XSPEC. To model the absorption of the Galactic halo we used a photoelectric absorption model in XSPEC, namely \phabs. The value of the foreground hydrogen absorption column was fixed at $6\times10^{20}$~cm$^{-2}$ based on the \citet{Dickey1990} HI maps, determined using the HEASARC $N_{\rm{H}}$ Tool\footnote{\burl{http://heasarc.gsfc.nasa.gov/cgi-bin/Tools/w3nh/w3nh.pl}}. An additional absorption component (\vphabs) was added for the power law component to account for the absorption of the background cosmological sources by material in the LMC. The abundances of this component were fixed to LMC values. 

\par
\begin{figure*}
\begin{center}
\resizebox{\hsize}{!}{\includegraphics[trim= 0.1cm 0.1cm 0.1cm 0.1cm, clip=true, angle=0]{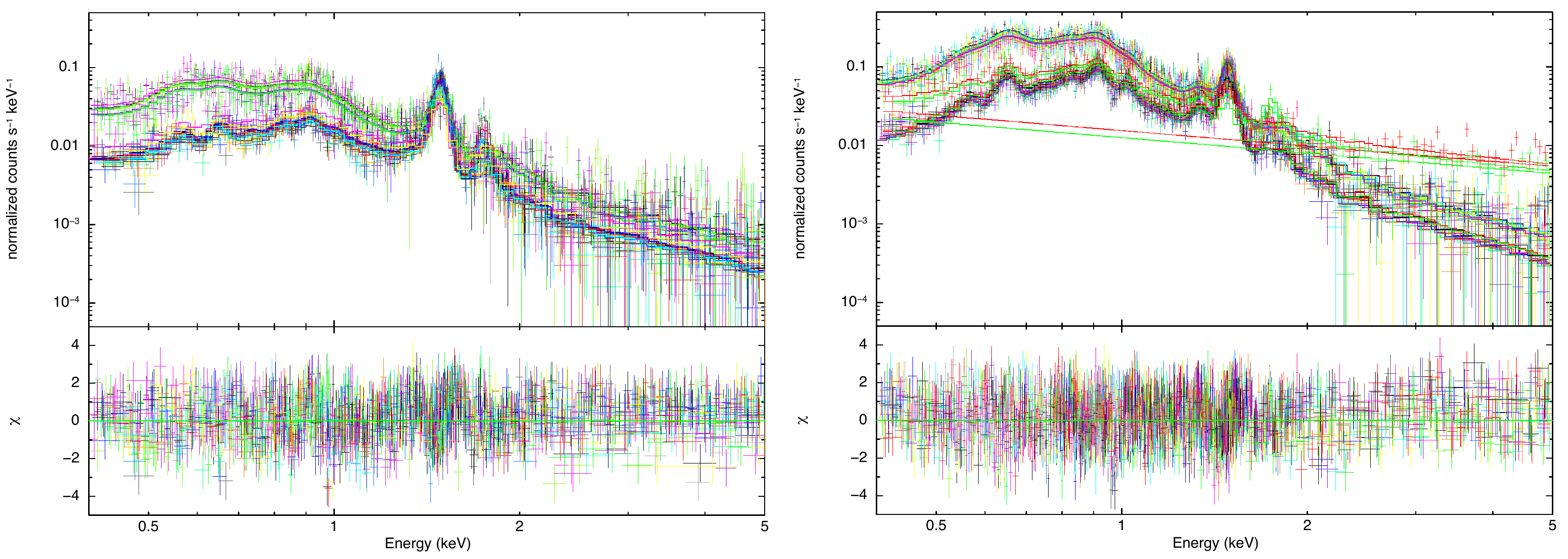}}
\caption{\textit{Left}: Simultaneous spectral fits to the EPIC BG-W spectra. \textit{Right}: Same as \textit{Left} for BG-E. The green and red straight lines represent the residual SP contamination in Obs.~ID~0113020201, in which only the EPIC-MOS1 and EPIC-MOS2 cameras were on. 
}
\label{bg-spectra}
\end{center}
\end{figure*}

 In principle, the AXBs of BG-E and BG-W should be of the same surface brightness as the components do not vary on such small spatial scales. We began by fitting the spectrum of BG-W which has lower surface brightness and tested if the increased absorption due to the foreground molecular cloud allows us to constrain the soft AXB emission, akin to a shadowing measurement. A simple fit with the normal AXB components was insufficient to adequately model the spectra. Thus, an additional thermal component (\vapec\ with LMC abundances) was included representing LMC ISM emission. This yielded a much improved fit with reduced $\chi^{2}$ ($\chi^{2}_{\nu} = 1.21$). The best-fit spectra are shown in Fig. \ref{bg-spectra}~($left$) with the fit results presented in Table \ref{bg_w_tab}. 

\begin{table}[htdp]
\caption{Spectral fit results for BG-W. See text for description of the model components.}
\begin{center}
\label{bg_w_tab}
\begin{tabular}{lll}
\hline
\hline
Component & Parameter  & Value  \\
\hline
\multicolumn{3}{c}{Foreground Absorption}\\
\hline
Galactic (\texttt{phabs}) & $N_{\rm{H,Gal}}$ ($10^{22}$ cm$^{-2}$) & 0.06 (fixed)\tablefootmark{a} \\
LMC (\vphabs)\tablefootmark{b} & $N_{\rm{H,LMC}}$ ($10^{22}$ cm$^{-2}$) & 0.87 (0.81--0.96)\tablefootmark{e} \\
& & \\
\hline
\multicolumn{3}{c}{Astrophysical background}\\
\hline
Local bubble (\vapec) & $kT$ (keV) & 0.1 (fixed)\tablefootmark{c}  \\
& $norm$ ($10^{-5}$ cm$^{-5}$) & 1.00 ($<2.39$) \\
Galactic halo (\vapec) & $kT$ (keV) & 0.23 (0.22--0.24) \\
& $norm$ ($10^{-5}$ cm$^{-5}$) & 6.50 (6.02--7.00) \\
Bkg. AGN (\pow) & $\Gamma$ & 1.46 (fixed) \\
& $norm$ ($10^{-5}$) & 1.28 (fixed)\tablefootmark{d} \\
& & \\
\hline
\multicolumn{3}{c}{LMC interstellar medium}\\
\hline
ISM (\vapec)\tablefootmark{b}  & $kT$ (keV) & 0.81 (0.80--0.84) \\
& $norm$ ($10^{-4}$ cm$^{-5}$) & 1.82 (1.60--2.05) \\
& & \\
\hline
Fit statistic & $\chi^{2}_{\nu}$ & 1.21 \\
\hline
\end{tabular}
\tablefoot{
\tablefoottext{a}{Fixed to the Galactic column density from the \citet{Dickey1990} HI maps.}
\tablefoottext{b}{Absorption and thermal component abundances fixed to those of the LMC.}
\tablefoottext{c}{$kT$ fixed to 0.1 keV, appropriate for the LHB emission (see text).}
\tablefoottext{d}{Normalisation fixed to equivalent of 10.5 photons~keV~cm$^{-2}$~s$^{-1}$~sr$^{-1}$ (see text).}
\tablefoottext{e}{Numbers in parentheses are the 90\% confidence intervals.}
}
\end{center}
\end{table}%

\par For the brighter BG-E region, we fixed the normal AXB contribution based on the BG-W results. With regard to the LMC ISM emission, we also kept the additional thermal component required in BG-W, fixing its temperature but allowing its normalisation to vary. The resulting fit yielded large residuals in the $0.4-0.7$~keV range, with $\chi^{2}_{\nu}~>2$. Thus, a second thermal component was added (\vapec\ with LMC abundances). While the fit was largely improved, there remained residuals at emission lines of some $\alpha$-group elements. Hence, we allowed O, Ne, and Mg abundances to vary and tied the abundances of the two thermal components representing the LMC ISM emission. This resulted in a substantially improved fit ($\chi^{2}_{\nu}=1.23$). The plasma temperatures of the two LMC ISM components are consistent with the ISM in other star-forming galaxies \citep{Mineo2012}. The best-fit spectra are shown in Fig. \ref{bg-spectra}~($right$) with the fit results presented in Table \ref{bg_e_tab}. 

\begin{table}[htdp]
\caption{Spectral fit results for BG-E. See text for description of the model components.}
\begin{center}
\label{bg_e_tab}
\begin{tabular}{lll}
\hline
\hline
Component & Parameter  & Value  \\
\hline
\multicolumn{3}{c}{Foreground Absorption}\\
\hline
Galactic (\texttt{phabs}) & $N_{\rm{H,Gal}}$ ($10^{22}$ cm$^{-2}$) & 0.06 (fixed)\tablefootmark{a} \\
LMC (\vphabs)\tablefootmark{b} & $N_{\rm{H,LMC}}$ ($10^{22}$ cm$^{-2}$) & 0.59 (0.55--0.64)\tablefootmark{e} \\
& & \\
\hline
\multicolumn{3}{c}{LMC interstellar medium}\\
\hline
ISM 1 (\vapec)\tablefootmark{b} & $kT$ & 0.23 (0.22--0.24) \\
& $norm$ ($10^{-4}$ cm$^{-5}$) & 8.93 (6.86--11.48) \\
ISM 2 (\vapec)\tablefootmark{c} & $kT$ & 0.81 (fixed) \tablefootmark{d}\\
& $norm$ ($10^{-4}$ cm$^{-5}$) & 4.13 (3.88--4.39) \\
& O ($Z/\rm{Z}_{\sun}$) & 1.25 (1.11--1.41) \\
& Ne ($Z/\rm{Z}_{\sun}$) & 1.26 (1.13--1.39) \\
& Mg ($Z/\rm{Z}_{\sun}$) & 1.34 (1.20--1.48) \\
& & \\
\hline
Fit statistic & $\chi^{2}_{\nu}$ & 1.23 \\
\hline
\end{tabular}
\tablefoot{
The normal astrophysical background parameters are fixed to the best-fit values determined from BG-W (see Table \ref{bg_w_tab}).\\
\tablefoottext{a}{Fixed to the Galactic column density from the \citet{Dickey1990} HI maps.}
\tablefoottext{b}{Absorption and thermal component abundances fixed to those of the LMC.}
\tablefoottext{c}{Only O, Ne, and Mg allowed to vary. All other metal abundances fixed to LMC values.}
\tablefoottext{d}{Fixed according to the results of the BG-W fits (see Table \ref{bg_w_tab}).}
\tablefoottext{e}{Numbers in parentheses are the 90\% confidence intervals.}
}
\end{center}
\end{table}%

\par From the fits to the background regions it is obvious that there is very significant LMC ISM emission in the \SB\ region. While we attempted to fit this emission with physically motivated models, a truly detailed analysis and interpretation of the ISM emission is beyond the scope of this paper. Such a study of the hot gas in the LMC will be presented in a future work. For our purposes, the best-fit models to the BG-E and BG-W spectra were simply fixed in the fits to the \SB\ spectra. We briefly note however that other interpretations are possible, e.g., the two-component LMC ISM emission could dominate the foreground Galactic emission, and only the ISM components are required in the fit.



\subsubsection{Analysis regions}
\label{an-regions}
The substantial \xmm\ data allowed us to perform a spatially resolved spectral study of \SB, much in the same way as BU04 only on smaller spatial scales. In this way we can analyse the variation in spectral components in different regions of the remnant in unprecedented detail. We follow the lead of BU04 with the nomenclature of our analysis regions, approximately splitting the superbubble shell into SE, NE, NW, and SW quadrants which are labelled shells A, B, C, and D, respectively. These shells are further subdivided into regions of interest, selected due to notable features in the three-colour X-ray image (see Fig. \ref{xmm-rgb}~$right$). Shell~A is the region of the brightest soft X-ray emission and, for the first time, we have resolved the brightest soft emission in shell~A, due to \SNR, from the thermal superbubble emission. For consistency in nomenclature, we label the SNR region and the bright superbubble emission to the north A1 and A2, respectively. Shell B contains part of the non-thermal shell. The deep \xmm\ image reveals significant structure in the non-thermal emitting regions, with two bright regions of the outer shell and a fainter filament slightly inside these. We label these B1 (SE of outer shell), B2 (NE of outer shell), and B3 (inner filament). Shell C contains the brightest part of the non-thermal shell. As with shell B, much detail is evident in the morphology of the hard emission. We subdivide this shell into C1 (the brightest region to the north) and C2 (the fainter region to the SW). Finally, Shell D, which is connected to Shell C by a thin hard filament, cannot be subdivided and is taken as a single complete region. In addition to the bright shell regions, the data also afford us the opportunity to study the fainter interior regions of \SB. We label these the I-regions which are defined as follows: I1, located inside shells A and B, and contains bright soft emission; I2, taken as the centre of the superbubble, this region is free of any limb-brightened shell emission; and I3, located inside shells C and D, whose hard X-ray structure is more diffuse than the shell. 


\subsubsection{Spectral fits}
\label{spec-fits}
\begin{figure*}
\begin{center}
\resizebox{\hsize}{!}{\includegraphics[trim= 0cm 0cm 0cm 0cm, clip=true, angle=0]{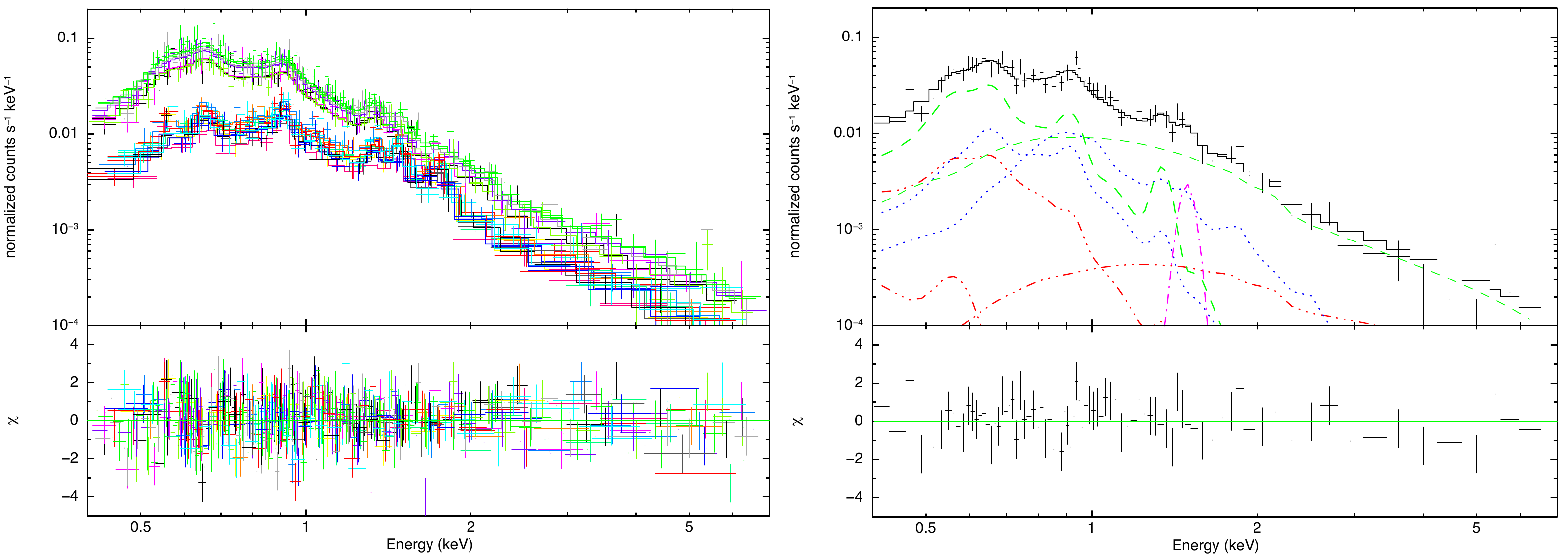}}
\caption{\textit{Left}: Simultaneous spectral fits to the EPIC A2 spectra. \textit{Right}: EPIC-pn spectrum of A2 from Obs.~ID~ 0601200101 (deepest EPIC-pn observation) with additive model components shown. The red dash-dot-dot-dot lines represent the AXB components, the magenta dash-dot line shows the instrumental fluorescence line, blue dotted lines mark the LMC ISM, and green dashed lines represent the source components (\texttt{vapec+pow}). Best-fit parameters are given in Table~\ref{fit-results}. 
}
\label{a2-spectra}
\end{center}
\end{figure*}

\textbf{Shell A (SE):} Due to its added complexity, the description of the spectral fits to A1 (\SNR) are deferred to Section \ref{snr-fit}. Region A2 is a little more distinct than other regions of superbubble thermal emission (see B1 and I1 below) as it is brighter and contains a filamentary structure. Soft emission lines in its spectra cannot be explained by the thermal background components alone. In addition, a hard tail is present which is most likely non-thermal in origin. Hence, we fit the spectra using a thermal plasma (\vapec\ with LMC abundance) plus power law model. Due to the relatively low number of hard photons, it proved difficult to constrain the slope of the power law component if left free. Thus, we fixed the slope of the power law to $\Gamma = 2.55$, which is the average of the slopes determined for the adjacent I1 and I2 regions (see below). The resulting fits yielded $\chi^{2}_{\nu} \sim 1.3$. Obvious residuals at $\sim0.7$~keV (O~VIII), $\sim0.9$~keV (Ne~IX), and $\sim1.4$~keV (Mg~XI) remained. Thus, we freed the abundances of these elements, resulting in an improved fit with $\chi^{2}_{\nu} = 1.08$. The best fit model parameters are given in Table \ref{fit-results} with the spectra shown in Fig. \ref{a2-spectra}. \\

\par \noindent \textbf{Shell B (NE):} B1 contains relatively bright hard emission as well as part of the enhanced soft emission in the east of \SB. This is evident in the spectra as obvious emission lines at $\sim0.7$~keV (O~VIII), $\sim0.9$~keV (Ne~IX), and $\sim1.4$~keV (Mg~XI). Hence, we fitted the B1 spectra with a thermal plasma with LMC abundance  (\vapec) plus a power law model. While the resulting fit statistic ($\chi^{2}_{\nu} = 1.11$) is relatively good, we suspected that this could be improved further as residuals remained at the aforementioned emission lines as well as below 0.5 keV. Hence, we allowed the abundances of O, Ne, and Mg to vary while fixing the remaining elemental abundances to LMC values. The fit was further improved ($\chi^{2}_{\nu} = 0.96$). The best fit model parameters are given in Table \ref{fit-results} with the spectra shown in Fig. \ref{b1-spectra}\footnote{The remaining spectral fit figures are deferred to Appendix \ref{app-fig}}. 

\par In contrast to B1, the spectrum of B2 is comparatively featureless, which is unsurprising considering this region is the second brightest in the 2--7 keV band so the hard emission dominates. We fitted the B2 spectra with a power law on top of the background components and found an acceptable fit ($\chi^{2}_{\nu} = 1.03$). Adding thermal plasma models did not improve the fits and were difficult to constrain. The best fit model parameters are given in Table \ref{fit-results} with the spectra shown in Fig. \ref{b2-spectra}. 

\par Due to the number of counts being $<1000$ for all of the EPIC-MOS spectra, only the B3 EPIC-pn spectra were used for the fits. B3 contains a hard filamentary structure so a non-thermal component was expected in the spectra. This was indeed evident as a hard tail. While the spectra of B3 exhibit some emission lines, preliminary fitting showed that these were most likely due to the background thermal components. We fitted the B3 spectra with a power law model on top of the background. The resulting fit was relatively poor with a $\chi^{2}_{\nu} = 1.33$. Attempts to improve the model by introducing additional thermal components were unsuccessful, resulting in non-sensical values for the fit parameters.  The best fit model parameters are given in Table \ref{fit-results} with the spectra shown in Fig. \ref{b3-spectra}. \\

\par \noindent \textbf{Shell C (NW):} The Shell C spectrum is rather featureless, indicative of a non-thermal origin. Thus, on top of the background model we added a power law to fit each of the subdivided C shells. The resulting fits to the C1 and C2 sub-regions were acceptable (see Table~\ref{fit-results}), with $\chi^{2}_{\nu} = 1.05$ and $1.06$, respectively. The best fit spectra of C1 and C2 are shown in Fig. \ref{c1-spectra} and Fig. \ref{c2-spectra}, respectively, with results in Table~\ref{fit-results}. YB09 reported that a simple power law fit to their `West' spectrum, equivalent to our C shell, could be statistically rejected in favour of a broken power law or with the \srcut\ model in XSPEC which represents a synchrotron spectrum from an exponentially cut off power-law distribution of electrons \citep{Reynolds1998}. Thus, we also fit the brightest region of shell C, namely C1, with an \srcut\ model in XSPEC. We follow the lead of YB09 and fixed the spectral index at 1~GHz to a range of values typical of SNRs, $\alpha = -(0.4-0.6)$. Our fits yielded a spectral roll-off frequency in the range $(1.5-3.5) \times 10^{17}$~Hz, similar to the results of YB09. However, the resulting $\chi^{2}_{\nu} \sim 1.09$, is not a statistical improvement on a simple power law fit. We applied the \srcut\ models to other regions around \SB\ with similar results, i.e., we cannot reject either the simple power law or the \srcut\ models based on our X-ray spectral fits alone. This is most likely due to the lower upper-limit to the energy range in our fits ($7$~keV), with YB09 fitting up to $>10$~keV. The physical implications of this model are discussed in Section~\ref{nt-emission}. \\

\par \noindent \textbf{Shell D (SW):} Similar to shell C, the spectrum of shell D is relatively featureless. We fitted this region with a power law in addition to the background, which yielded a  $\chi^{2}_{\nu} = 1.01$. The best fit model parameters are given in Table \ref{fit-results} with the spectra shown in Fig. \ref{d-spectra}. \\

\par \noindent \textbf{Interior regions:} I1 is substantially different to regions I2 and I3 in that there is very obvious bright thermal emission present. This is evident in the spectra as obvious emission lines at $\sim0.7$~keV (O~VIII), $\sim0.9$~keV (Ne~IX), $\sim1.0$~keV (Ne~X), and $\sim1.4$~keV (Mg~XI). In addition, there is a high energy tail and, thus, a non-thermal component may also be present. Motivated by these features, we fitted the I1 spectra with a thermal plasma with LMC abundance (\vapec) plus a power law model. The resulting fit, with $\chi^{2}_{\nu} = 1.33$, failed to properly account for emission lines at 0.9~keV, 1~keV, and 1.4~keV, as well as yielding a photon index of $\sim3.4$, inconsistent with the non-thermal emission from adjacent regions of \SB. Thus, we allowed the abundances of O, Ne, and Mg to vary while fixing the remaining elemental abundances to LMC values. This improved the fit to $\chi^{2}_{\nu} =1.06$ and resulted in a more reasonable photon index of $\Gamma=2.59$. The best fit model parameters are given in Table \ref{fit-results} with the spectra shown in Fig. \ref{i1-spectra}. 
\par No thermal component was required for I2 and I3 with a power law component on top of the astrophysical background sufficient  for acceptable fits ($\chi^{2}_{\nu} = 1.03$ and 1.06, respectively). The best fit model parameters are given in Table \ref{fit-results} with the spectra shown in Figs. \ref{i2-spectra} and \ref{i3-spectra}.

\afterpage{
\begin{landscape}
\vspace*{\fill}
\begin{table}[h]
\centering
\caption{Results of simultaneous spectral fits to \SB\ regions}
\begin{center}
\label{fit-results}
\begin{tabular}{cccccccccc}
\hline
 & & & &  & & & &  &   \\
 &  LMC absorption & \multicolumn{2}{c}{Non-thermal (\pow)}  & \multicolumn{5}{c}{Thermal (\vapec)} &  \\
 &  \hrulefill  & \multicolumn{2}{c}{\hrulefill}  & \multicolumn{5}{c}{\hrulefill} &  \\
 Region  & $N_{\rm{H,LMC}}$  & $\Gamma$ & $norm$ & $kT$ & O & Ne & Mg & $norm$ & $\chi^{2}_{\nu}$\\
 & ($10^{22}$ cm$^{2}$) & & ($10^{-4}$)  & (keV) & ($Z/\rm{Z}_{\sun}$) & ($Z/\rm{Z}_{\sun}$) & ($Z/\rm{Z}_{\sun}$) & ($10^{-4}$ cm$^{-5}$) &  \\ 
\hline
\hline
A2 & $0.52~(0.45-0.59)$ & $2.55~(fixed)$ & $0.21~(0.20-0.22)$ & $0.18~(0.17-0.19)$ & $1.04~(0.72-1.72)$ & $1.31~(0.88-2.24)$ & $>4.39$ & $1.92~(0.83-3.94)$  & 1.08    \\
 & & & &  & &  & & &   \\
 & & & &  & &  & & &   \\
B1 & $0.43~(0.40-0.46)$ & $2.73~(2.65-2.81)$ & $0.57~(0.53-0.61)$ & $0.31~(0.28-0.34)$ & $2.96~(1.30-5.25)$ & $3.29~(2.26-5.35)$ & $6.04~(3.94-9.75)$ & $0.26~(0.16-0.40)$  & 0.96    \\
B2 & $0.54~(0.52-0.55)$ & $2.50~(2.47-2.53)$ & $1.29~(1.26-1.32)$ & -- & -- & -- & -- & --  & 1.03    \\
B3 & $0.51~(0.48-0.54)$ & $2.75~(2.62-2.88)$ & $0.29~(0.24-0.29)$ & -- & -- & -- & -- & --  & 1.33    \\
 & & & &  & &  & & &   \\
C1 & $1.00~(0.97-1.02)$ & $2.32~(2.29-2.35)$ & $1.64~(1.57-1.70)$ & -- & -- & -- & -- & -- & 1.05    \\
C2 & $1.38~(1.33-1.42)$ & $2.42~(2.38-2.46)$ & $1.02~(0.97-1.05)$ & -- & -- & -- & -- & -- & 1.06    \\
 & & & &  & & &  & &   \\
D & $0.65~(0.62-0.67)$ & $2.45~(2.40-2.49)$ & $0.50~(0.48-0.52)$ & -- & -- & -- & -- & -- & 1.01    \\
 & & & &  & &  & & &   \\
I1 & $0.37~(0.36-0.39)$ & $2.59~(2.49-2.68)$ & $0.44~(0.41-0.48)$ & $0.40~(0.37-0.46)$ & $3.34~(1.91-5.19)$ & $4.90~(3.86-6.12)$ & $>9.16$ & $0.18~(0.15-0.23)$ & 1.06    \\
I2 & $0.73~(0.71-0.75)$ & $2.51~(2.45-2.56)$ & $0.61~(0.57-0.65)$ & -- & -- & -- & -- & -- & 1.03    \\
I3 & $1.01~(0.96-1.04)$ & $2.34~(2.31-2.38)$ & $1.31~(1.25-1.36)$ & -- & -- & -- & -- & -- & 1.06    \\
 & & & &  & & & & &   \\
\hline
\multicolumn{10}{l}{The upper and lower limits correspond to the 90\% confidence intervals of the fit parameters.}
\end{tabular}
\end{center}
\end{table}%
\vspace*{\fill}
\end{landscape}
}

\subsubsection{A1 spectral fits: \SNR}
\label{snr-fit}
The spectra of A1 show clear emission lines at $\sim0.6$~keV (O~VII), $\sim0.7$~keV (O~VIII), $\sim0.9$~keV (Ne~IX), $\sim1.0$~keV (Ne~X), $\sim1.4$~keV (Mg~XI), and $\sim1.84$~keV (Si~XIII), indicative of a thermal plasma with a temperature of $10^{6-7}$~K. Fitting the spectrum is not as straight forward as in the other regions as the shell in A1 is immersed in contaminating emission from \SB. To account for this, we assume that the SB emission from the adjacent region I1 is representative of the contaminating emission in A1, and include this component in the models accordingly. We first attempted to fit the shell emission using a simple \vpshock\ model with LMC abundance of $0.5~Z/\rm{Z}_{\sun}$ \citep{Russell1992}. This failed to adequately account for the strong emission lines of O, Ne, Mg, and Si. As such we allowed the abundances of these elements to vary while keeping the remaining metals fixed to the LMC value. This yielded an acceptable fit with $\chi^{2}_{\nu} = 1.19$, the results of which are given in Table \ref{a1_tab}. However, this model did not provide strong constraints on the abundance parameters. For this simple model, the derived plasma temperature is higher than in the surrounding regions, with large overabundance of $\alpha$-process elements. This is consistent with an SNR origin, likely in the transition between free-expansion and the Sedov phase. 

\par
\begin{figure*}[!ht]
\begin{center}
\resizebox{\hsize}{!}{\includegraphics[trim= 0cm 0cm 0cm 0cm, clip=true, angle=0]{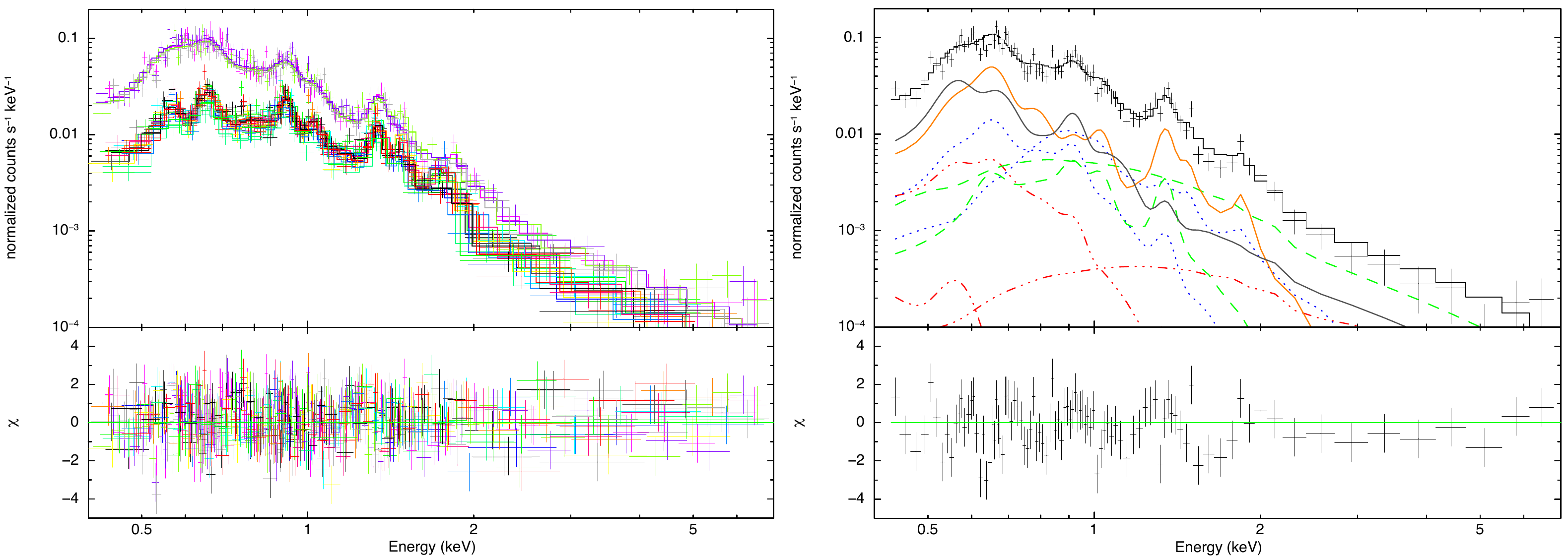}}
\caption{\textit{Left}: Simultaneous spectral fits to the EPIC spectra of \SNR. \textit{Right}: EPIC-pn spectrum of \SNR\ from Obs.~ID~0601200101 with additive model components shown. The red dash-dot-dot-dot lines represent the AXB components, blue dotted lines mark the LMC ISM, and green dashed lines represent the contamination from \SB. The black solid line represents the swept-up ISM component and the ejecta emission in shown by the orange solid line. Best-fit parameters are given in Table~\ref{a1_tab}. 
}
\label{a1-spectra}
\end{center}
\end{figure*}

 Assuming the X-ray emission from this SNR arises from the combination of an ejecta dominated and an ISM dominated shock, then a more representative model would consist of two thermal plasma components. We must be cautious however as fitting a multi-component SNR model, in addition to the background \SB\ components could lead to problems in the fit, namely, the contributions to the continuum are difficult to constrain. Thus, a simple, though physically plausible model is required. We assume a pure metal plasma consisting of O, Ne, Mg, and Si for the ejecta and an additional component representing the swept-up ISM shocked by the blast wave. This ISM component is likely more significant in the north of \SNR\ where the soft emission is brightest. Hence, we fit the spectra with a \vpshock+\vpshock\ model. The ISM component has abundances fixed to the LMC values. For the ejecta component, we follow the method of \citet{Vink1996}, which allows us to fit the ejecta with a simple model, but can provide detailed information on the abundance ratios in the ejecta. We assume that the ejecta consist mainly of O, fix the O abundance at a large value ($10^{4}~Z/\rm{Z}_{\sun}$), and allow the abundances of Ne, Mg, and Si to vary relative to it. All other abundances were fixed to 0. This model is still oversimplified. We must assume that the ejecta components have the same temperature and ionisation conditions. However, given the already complex model, the addition of individual pure-metal plasma components for each element exacerbates the situation and strong constraints on the model parameters cannot be obtained. Hence, we continued with the simplifying assumption of a uniform ejecta temperature. The resulting fit was acceptable fit with $\chi^{2}_{\nu} = 1.16$, the results of which are given in Table \ref{a1_tab} and the spectra shown in Fig. \ref{a1-spectra}. These results are discussed in detail in Section \ref{new-snr}.

\begin{table}
\caption{Spectral fit results for A1 (\SNR). See text for description of the models.}
\begin{center}
\label{a1_tab}
\begin{tabular}{llr}
\hline
Component & Parameter & Value\\
\hline
\hline
\multicolumn{3}{c}{Model: simple \vpshock}\\
\hline
\vphabs & $N_{\rm{H,LMC}}$ ($10^{22}$ cm$^{-2}$) &   0.36 (0.32--0.40)\tablefootmark{a,d}  \\
\vpshock \tablefootmark{b}  & $kT$ & 3.61 (3.32--3.83) \\
 & O ($Z/\rm{Z}_{\sun}$) & 7.20 (4.41--8.63) \\
 & Ne ($Z/\rm{Z}_{\sun}$) & 4.73 (2.70--5.50) \\
 & Mg ($Z/\rm{Z}_{\sun}$) &  9.89 ($>6.07$) \\
 & Si ($Z/\rm{Z}_{\sun}$) & 8.16 ($>4.82$) \\
 & $\tau_{u}$ ($10^{10}$~s~cm$^{-3}$) & 1.77 (1.56--2.05) \\
 & $norm$ ($10^{-6}$ cm$^{-5}$) & 1.84 (1.64--3.13) \\
\multicolumn{3}{c}{ }\\
Fit statistic & $\chi^{2}_{\nu}$ & 1.19 \\
\multicolumn{3}{c}{ }\\
\hline
\hline
\multicolumn{3}{c}{Model: ejecta + ISM}\\
\hline
\vphabs & $N_{\rm{H,LMC}}$ ($10^{22}$ cm$^{-2}$) &   0.46 (0.41--0.50)\tablefootmark{a,d}  \\
\vpshock\ $(ejecta)$ & $kT$ & 4.09 (3.64--4.41) \\
 & O ($10^{4}~Z/\rm{Z}_{\sun}$) & 1.00 (fixed) \\
 & Ne ($10^{4}~Z/\rm{Z}_{\sun}$) & 0.38 (0.31--0.48) \\
 & Mg ($10^{4}~Z/\rm{Z}_{\sun}$) & 0.71 (0.56--0.93) \\
 & Si ($10^{4}~Z/\rm{Z}_{\sun}$) & 0.27 (0.17--0.46) \\
 & $\tau_{u}$ ($10^{10}$~s~cm$^{-3}$) & 9.70 (6.00--11.40) \\
 & $norm$ ($10^{-9}$ cm$^{-5}$) & 3.96 (2.36--5.70) \\
 & $L_{X}$\tablefootmark{c} ($10^{34}$ erg~s$^{-1}$) & 3.3 \\
\multicolumn{3}{c}{ }\\
\vpshock\ $(ISM)$ & $kT$ & 2.71 (1.70--3.81) \\
 & $\tau_{u}$ ($10^{10}$~s~cm$^{-3}$) & 0.59 (0.50--0.67) \\
 & $norm$ ($10^{-5}$ cm$^{-5}$) & 1.89 (1.39--2.48) \\
 & $L_{X}$\tablefootmark{c} ($10^{34}$ erg~s$^{-1}$) & 5.1 \\
\multicolumn{3}{c}{ }\\
Fit statistic & $\chi^{2}_{\nu}$ & 1.16 \\
\multicolumn{3}{c}{ }\\
\hline
\end{tabular}
\tablefoot{
\tablefoottext{a}{Absorption component abundances fixed to those of the LMC.}
\tablefoottext{b}{Only O, Ne, Mg, and Si allowed to vary. All other metal abundances fixed to the LMC value of $0.5~Z/\rm{Z}_{\sun}$ \citep{Russell1992}.}
\tablefoottext{c}{De-absorbed 0.3-10~keV X-ray luminosity, adopting a distance of 50~kpc to the LMC.}
\tablefoottext{d}{The numbers in parentheses are the 90\% confidence intervals.}
}
\end{center}
\end{table}%

\subsection{Radio}
\label{radio-analysis}
\begin{figure}[!h]
\resizebox{\hsize}{!}{\includegraphics[trim=0cm 2cm 1.7cm 1.7cm, clip=true, angle=0]{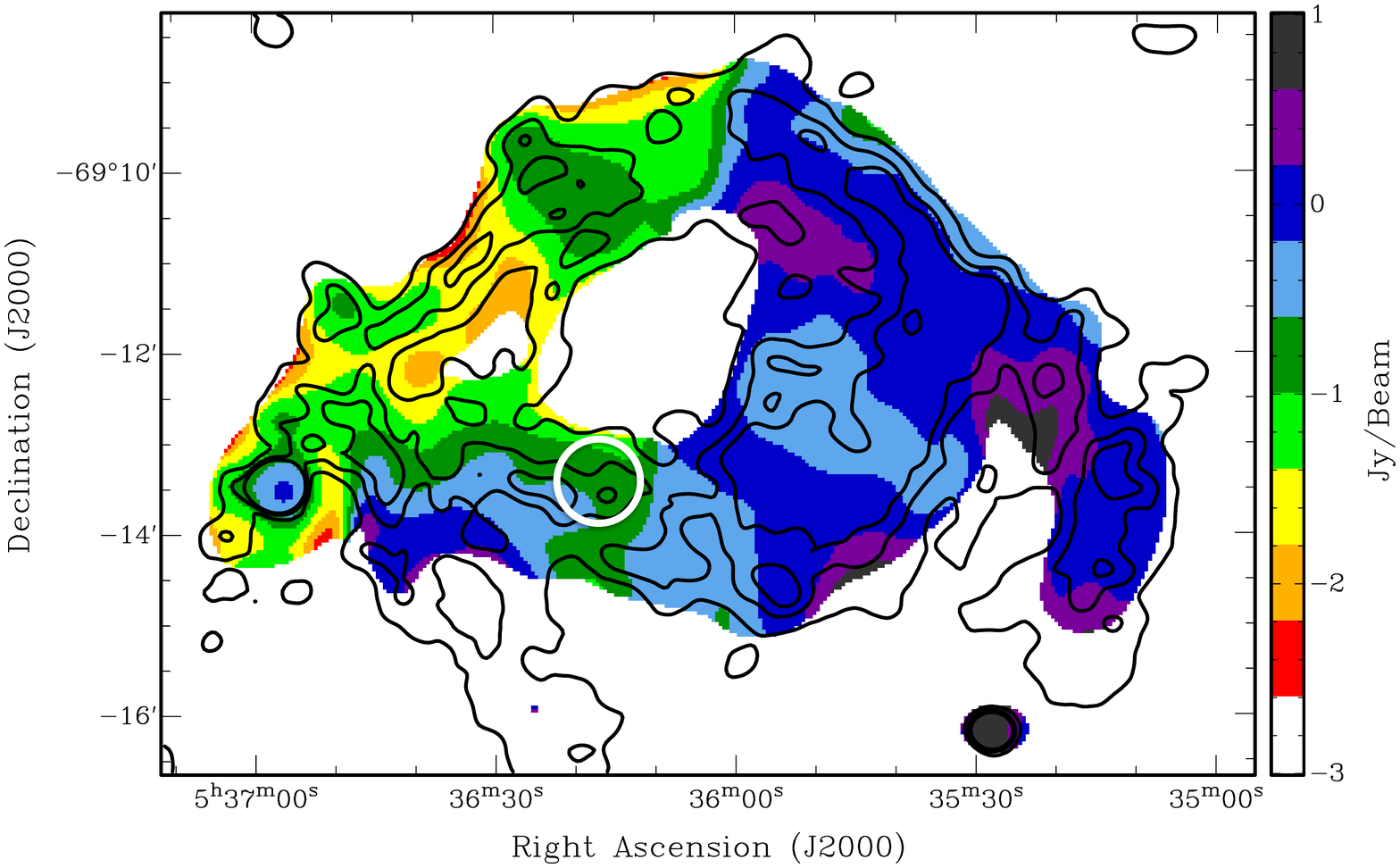}}
\caption{Radio-continuum spectral map of \SB\ between 36~cm and 20~cm. 20~cm contours have been superimposed at levels of 1, 3 and 5 mJy. The white circle in the southern shell shows the approximate extent of the SNR. The sidebar on the right quantifies the radio spectral index.
\label{rc-spcmp}}
\end{figure}

Spatial distribution of the spectral indices can be seen in Fig. \ref{rc-spcmp}, where the change in flux density is shown across \SB\ between wavelengths of 36~cm and 20~cm. We produced this image by reprocessing all observations to a common $u - v$ range, and then fitting $S \propto\ \nu^{\alpha}$ pixel by pixel using both images simultaneously. This image shows a distinct variation between the western and eastern sides of \SB, where the eastern side shows steeper spectral indices ($-2.0 < \alpha < -0.5$), while the western side shows much flatter ($-0.5 < \alpha < 0.5$), indicative of thermal emission. The majority of the SNR exhibits a spectral index of $\sim-0.7$, which is consistent with younger SNRs \citep[examples given in][]{Bozzetto2014}. However, this value is not only constrained to the immediate vicinity of the remnant, and such values can be seen extending well beyond the extent of the SNR in both the eastern and southern directions.

\begin{figure}[!h]
\resizebox{\hsize}{!}{\includegraphics[trim= 2cm 2.5cm 1.5cm 2cm, clip=true, angle=-90]{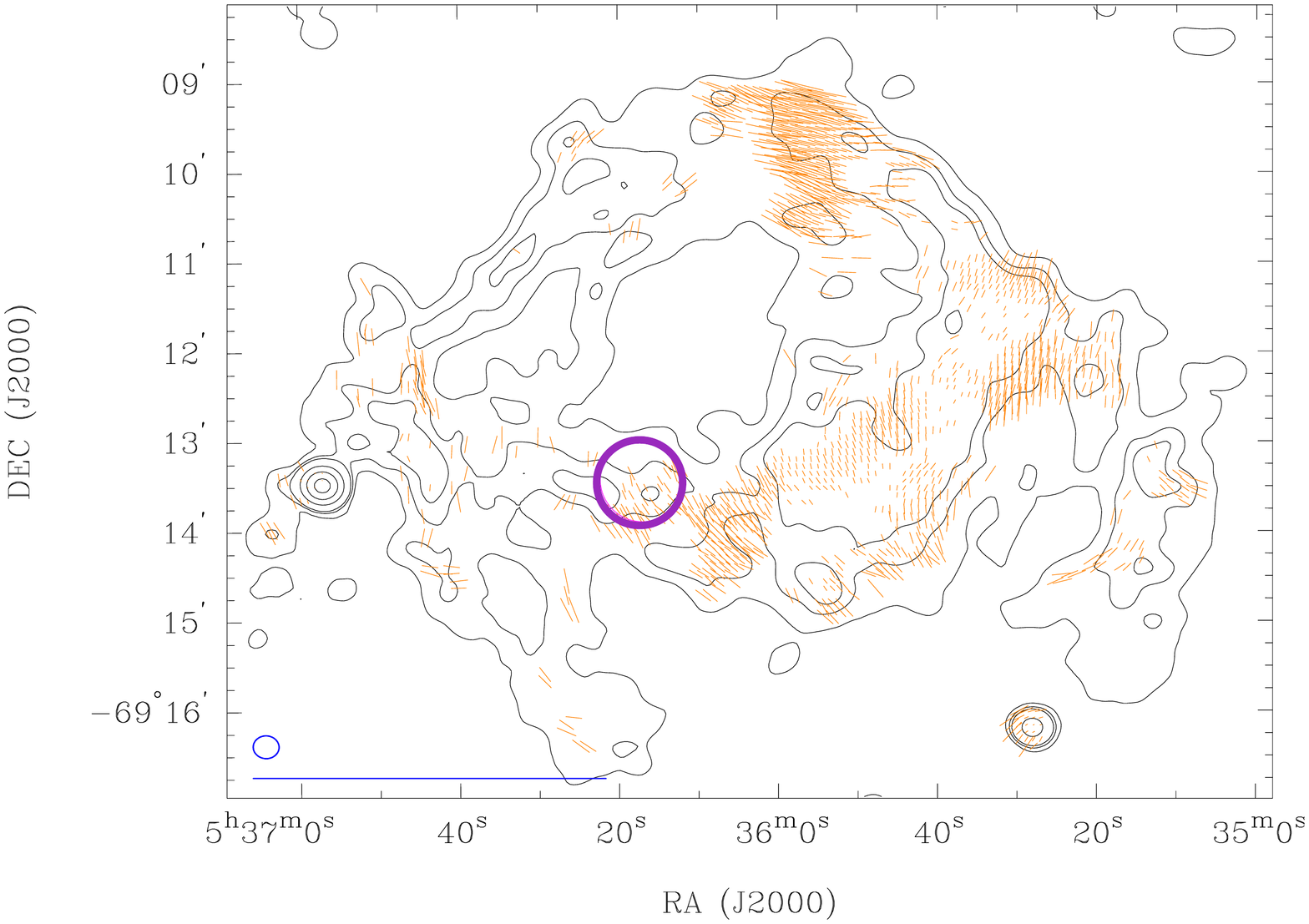}}
\caption{20~cm magnetic-field vectors overlaid on 20~cm contours (1, 3, 5, 20, 50 and 100 mJy) of \SB\ from ATCA observations. The magenta annotation shows the approximate extent of the SNR. The ellipse in the lower-left corner represents the synthesized beamwidth of  17.3\arcsec $\times$  15.2\arcsec, and the line below the ellipse represents a polarization vector of 100 per cent.
 \label{rc-pol}}
\end{figure}

We calculated the fractional polarisation {(\textit P)} at 20~cm using:

\begin{equation}
P=\frac{\sqrt{S_{Q}^{2}+S_{U}^{2}}}{S_{I}}
\end{equation}

\noindent where $S_{Q}, S_{U}$ and $S_{I}$ are integrated intensities for \textit{Q}, \textit{U} and \textit{I} Stokes parameters (Fig.~\ref{rc-pol}). We estimate a mean fractional polarisation of 4$\pm$1\% across the region of \SB.

\subsection{Optical}
\label{optical-analysis}
\begin{figure}[!h]
\begin{center}
\resizebox{\hsize}{!}{\includegraphics[trim= 0cm 0cm 0.7cm 0cm, clip=true, angle=0]{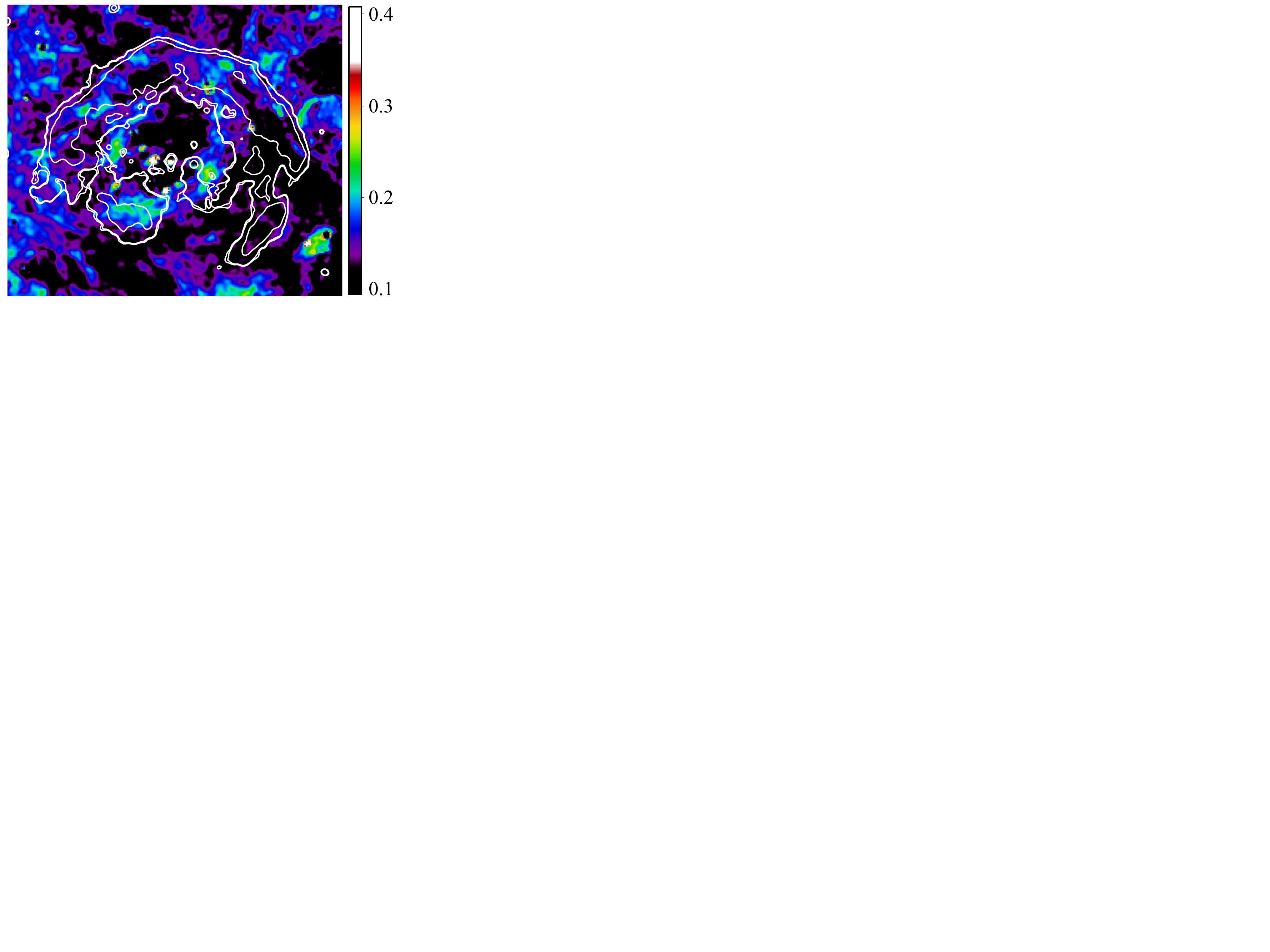}}
\caption{MCELS [\ion{S}{ii}]/H$\alpha$ ratio image of \SB\ with $1-2$~keV contours from Fig.~\ref{mcels-3pan}-middle. The colour scale indicates the value of [\ion{S}{ii}]/H$\alpha$.}
\label{ratio}
\end{center}
\end{figure}

\citet{Mathewson1985} reported on the analysis of the optical emission from \SB, including an analysis of the [\ion{S}{ii}]/H$\alpha$ ratio. An [\ion{S}{ii}]/H$\alpha$ flux ratio $>0.4$ is indicative of the presence of an SNR \citep{Mathewson1973,Fesen1985}. \citet{Mathewson1985} found that this ratio is $<0.3$ around the superbubble. We repeated the optical emission line analysis with the MCELS data (see Fig.~\ref{ratio}), the results of which are consistent with those of \citet{Mathewson1985}. The interpretation of the lack of optical emission from \SNR\ is discussed in Section~\ref{new-snr}.

\section{Discussion}
\label{dis}

\subsection{\SB\ multi-wavelength morphology}
\label{mwm}
\begin{figure*}
\begin{center}
\resizebox{\hsize}{!}{\includegraphics[trim= 0.3cm 16.3cm 4.5cm 0.3cm, clip=true, angle=0]{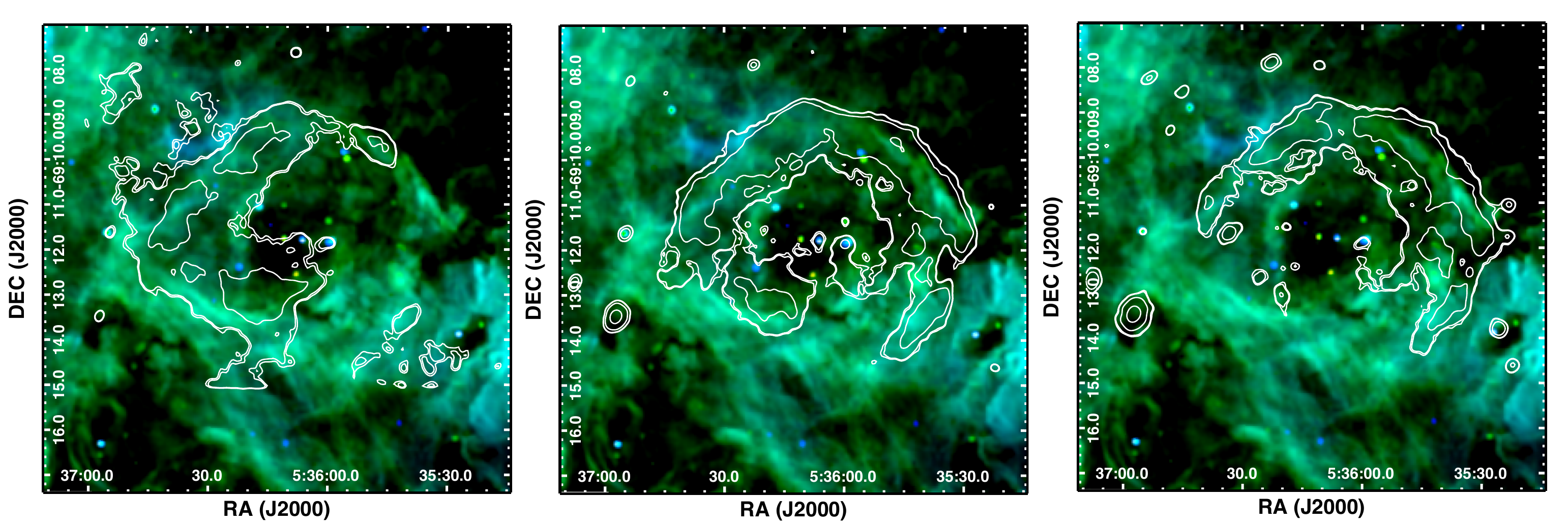}}
\caption{MCELS RGB ([\ion{S}{ii}], H$\alpha$, [\ion{O}{iii}]) image of \SB\ with $0.3-1$~keV contours (\textit{left}), $1-2$~keV contours (\textit{middle}), and $2-7$~keV contours (\textit{right}). The contour levels in each case were chosen arbitrarily to highlight X-ray features discussed in Section~\ref{an-regions}. It is clear from these images that the hard X-ray morphology is tightly correlated with the northern H$\alpha$ shell.}
\label{mcels-3pan}
\end{center}
\end{figure*}

The H$\alpha$ shell of \SB\ is very well defined. It confines the thermal superbubble emission in the east (see Fig.~\ref{mcels-3pan}~$left$) and correlates well with the non-thermal X-ray shell (see Fig.~\ref{mcels-3pan}~$right$). The 20~cm radio emission also follows very closely to the morphology of the H$\alpha$ shell. This is consistent with the standard superbubble picture of a pressure-driven bubble pushing out into the cool ISM with the photoionisation front due to the massive stellar population producing the H$\alpha$ shell. Indeed, the low [\ion{S}{ii}]/H$\alpha$ ratio ($<0.4$) throughout the superbubble \citep{Mathewson1985} points to photoionisation as the predominant mechanism producing the optical line emission. We have shown that the radio spectral indices around \SB\ are highly variable (Fig.~\ref{rc-spcmp}) with an obvious dichotomy between eastern and western shells. This has also previously been noted by \citet[][and references therein]{Mathewson1985}. The western shell exhibits very flat spectral indices ($0.5\gtrsim\alpha\gtrsim-0.5$), mostly consistent with a thermal origin. We interpret this flat spectral index as being due to contamination by the foreground molecular cloud covering the west of \SB. The eastern shell shows much steeper spectral indices ($-0.6\gtrsim\alpha\gtrsim-2.2$). The mean fractional polarisation at 20~cm is also quite low at 4$\pm$1\% across the region of \SB. The X-ray emission from \SB\ is also largely consistent with previous works in the literature. We discuss the properties of the X-ray emission in more detail in the forthcoming sections.



\subsection{Thermal X-ray emission}
\subsubsection{Superbubble}
\label{thermal-sb}
We detected thermal emission from the southeastern and eastern regions of the superbubble, as in the previous X-ray works on \SB. BU04 detected thermal emission from shell A, SW04 from the eastern half\footnote{We note that these authors also included a thermal component in their fits to the western half of \SB, however it was dominated by their non-thermal emission component.}, and YB09 from their SE and NE regions, approximately equivalent to shells A and B. Comparison of derived spectral parameters across the analysis is difficult due to the choice of analysis regions and, in particular, the identification of a new candidate SNR in shell A. However, it is clear in all cases that a soft thermal component ($kT < 1$~keV) with enhanced metal abundances is required to fit the thermal emission in \SB. 

\par We detected thermal emission in regions A1, B1, and I1, which effectively delineate the south eastern edge of \SB. This indicates a limb-brightened morphology which is consistent with the SB picture of thermal evaporation of cool material from the shell into the hot interior, as noted by SW04. There are several reasons why thermal emission was not detected from other regions of \SB. Firstly, the non-detection of thermal emission from shell C could be because of the higher absorbing column due to the foreground molecular cloud. In addition, any thermal components present in the spectra of shell C are likely dominated by the non-thermal emission, making their contribution to the spectra difficult to identify. We suspect that this may be the case for the regions B2, B3, and D where the absorption is less (by about half or lower) than in shell C, yet no thermal emission could be identified. However, in the case of B3 we note that the best-fit spectral model yielded a relatively poor fit ($\chi^{2}_{\nu} =1.33$) with residuals at thermal line energies, most notably at $\sim0.9$~keV (Ne IX). Attempts to improve the model by including a thermal component were unsuccessful, though it seems clear that there must be at least some thermal contribution. A further reason for the non-detection of thermal emission in other regions could simply be because the thermal emission is enhanced in the southeastern and eastern shell. This could be the case if an SN occurred near, and is now interacting with the shell wall, and/or the ambient density towards the east and southeast is higher.

\par The determined plasma temperatures in A2, B1, and I1 reveal slight variations between the regions. I1 exhibits the hottest thermal component with $kT = 0.40~(0.37 - 0.46)$, B1 being slightly cooler with $kT = 0.31~(0.28 - 0.34)$, and A2 being the coolest with $kT = 0.18~(0.17 - 0.19)$. Such plasma temperatures have been observed in many other LMC SBs \citep[e.g.,][]{Dunne2001,Cooper2004}. Another characteristic of the thermal emission is the overabundant O, Ne, and Mg to account for the observed line emission. Such $\alpha$-enrichment is evidence for a recent core-collapse (CC) SNR interaction with the shell, which also agrees with the suggestion that the thermal emission is enhanced in the southeastern and eastern regions due to an SNR impact on the shell. Additionally, the metal enrichment in this region can also result in a higher X-ray luminosity \citep{Silich2001}. The interaction of an off-centre SNR with an SB shell wall has also been suggested as an explanation for the overabundances and/or limb-brightened morphologies observed in LMC SBs DEM~L50 and DEM~L152 \citep{Jaskot2011}. 

\par Evidence for a recent SNR, responsible for the enhanced abundances in the east of the bubble, would be the presence of a compact object. BU04 detected three candidates in \SB. One of these, their Source 6, is located in the east of the superbubble, immersed in the thermal emission, and may be the compact remnant of the SN explosion responsible for the metal enrichment in the region. BU04 found a featureless spectrum for Source 6 which was best modelled with a power law with $\Gamma=1.8 (1.5-2.3)$ and the $0.5-9$~keV X-ray luminosity of $7.8\times10^{33}$~erg~s$^{-1}$. They also found no evidence for any long-term variability or pulsations from the object. To add to this analysis we extracted \xmm\ spectra from Source 6 taking the backgrounds from a nearby region in \SB\ to, as much as possible, account for the contaminating emission. Due to the low count rate of the object and the poorer resolution of \xmm\ we decided to merge the EPIC spectra from all the observations using the task \texttt{epicspeccombine}\footnote{The \texttt{epicspeccombine} task only became available in version 13 of the SAS. The observational data was re-processed accordingly to ensure compatibility with the newer version.}. Spectral fit results to the combined spectrum are fully consistent with those of BU06. We also do not observe any obvious long term variation in the flux of Source~6 though the low net counts in each observation makes this difficult to identify.

\subsubsection{\SNR}
\label{new-snr}
We obtained an acceptable fit for the X-ray spectrum of \SNR, assuming a physical model of ejecta plus swept-up ISM components (see Section \ref{snr-fit}). The detection of emission lines from $\alpha$-process elements in the ejecta component points to a CC origin for \SNR. It is possible to determine the abundance ratios in the ejecta based on the abundance parameters of the metals. These ratios can then be compared to theoretical explosive nucleosynthesis yield tables to determine the mass of the stellar progenitor. We assumed that the ejecta are well mixed, i.e., the abundance distribution in the shocked ejecta is representative of the ejecta in general, and it follows that the metals are co-spatial. We can estimate the emission measure of each metal from the normalisation parameter of the fit component. Even if the plasma comprises metals only, XSPEC outputs the normalisation ($K$) in terms of the emission measure of H ($n_{e}n_{\rm{H}}V$), i.e, 

\begin{equation}
\label{em}
K = \frac{10^{-14}}{4\pi D^{2}} n_{e}n_{\rm{H}}V .
\end{equation}

\begin{figure}[!h]
\begin{center}
\resizebox{\hsize}{!}{\includegraphics[trim= 0.1cm 6cm 6cm 0.1cm, clip=true, angle=0]{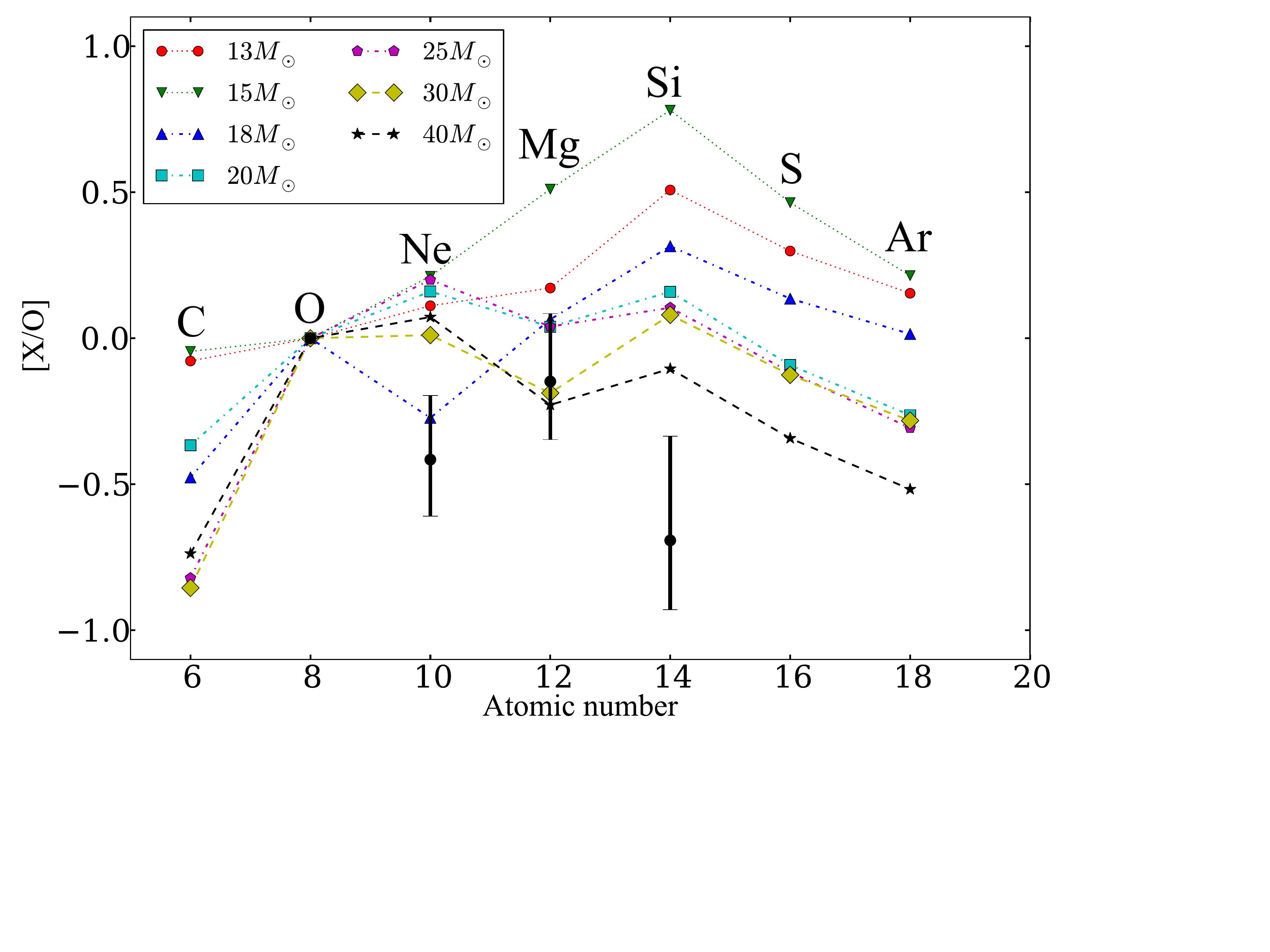}}
\caption{Abundance ratios of metals to O in the ejecta, relative to the corresponding solar ratios. \SNR\ data are in black. Theoretical yields for various progenitor masses determined from the Yields Table 2013 \citep[see also][]{Nomoto2013} are also included. Error bars are calculated from 90\% confidence intervals of fit parameters.
}
\label{yields}
\end{center}
\end{figure}

This equation can be adjusted to determine the emission measure ($n_{e}n_{\rm{X}}V$) for element $X$ by substituting $n_{\rm{H}}$ for $n_{X} =(n_{X}/n_{\rm{H}})_{\sun} (Z_{X}/Z_{X_{\sun}})$, where $(n_{X}/n_{\rm{H}})_{\sun}$ is the solar abundance of $X$ from \citet{Wilms2000} and $(Z_{X}/Z_{X_{\sun}})$ is the abundance of $X$ in the spectral fits. We determined the emission measure for O, Ne, Mg, and Si in this manner. Since we assume that the ejecta are well mixed, $n_{e}$ and $V$ can be taken as the same for each emission measure. Finally, we determined the value of [X/O] (the logarithm of the ratio of the X to the O abundance compared to the solar value) for Ne, Mg, and Si in order to compare the abundance ratios to the theoretical models. The value of [O/O] is naturally 0, and without error since we fixed the O abundance in our fits. For the comparison, we used data from the Yields Table 2013\footnote{Available at \burl{http://star.herts.ac.uk/\~chiaki/works/YIELD\_CK13.DAT}} \citep[see also][and references therein]{Nomoto2013}. In Fig.~\ref{yields} we show the theoretical explosive nucleosynthesis yields for a range of progenitor masses with approximately LMC metallicity,  adopting the canonical explosion energy of $10^{51}$~erg. The ejecta abundance pattern determined in our spectral analysis are also shown. The values of [Ne/O] and [Mg/O] suggest a progenitor mass of $\sim18$~M$_{\sun}$. [Si/O] is out of the range of progenitor masses. A possible reason for this is that, contrary to our earlier assumption, the observed Si abundance in the shocked ejecta is not representative of the ejecta as a whole. This might be the case if the higher mass elements are located closer to the interior of \SNR\ and are yet to be shocked by the reverse shock meaning the [Si/O] value in the outer ejecta is misleading. We also note that the error bars are determined from the 90\% confidence intervals of the fit parameters. Applying a more stringent error constraint results in larger error bars. Consequently, the data would be more consistent with a higher mass progenitor of $\gtrsim40$~M$_{\sun}$. 


\par
\begin{figure}[!h]
\begin{center}
\resizebox{\hsize}{!}{\includegraphics[trim= 0cm 0cm 0cm 0cm, clip=true, angle=0]{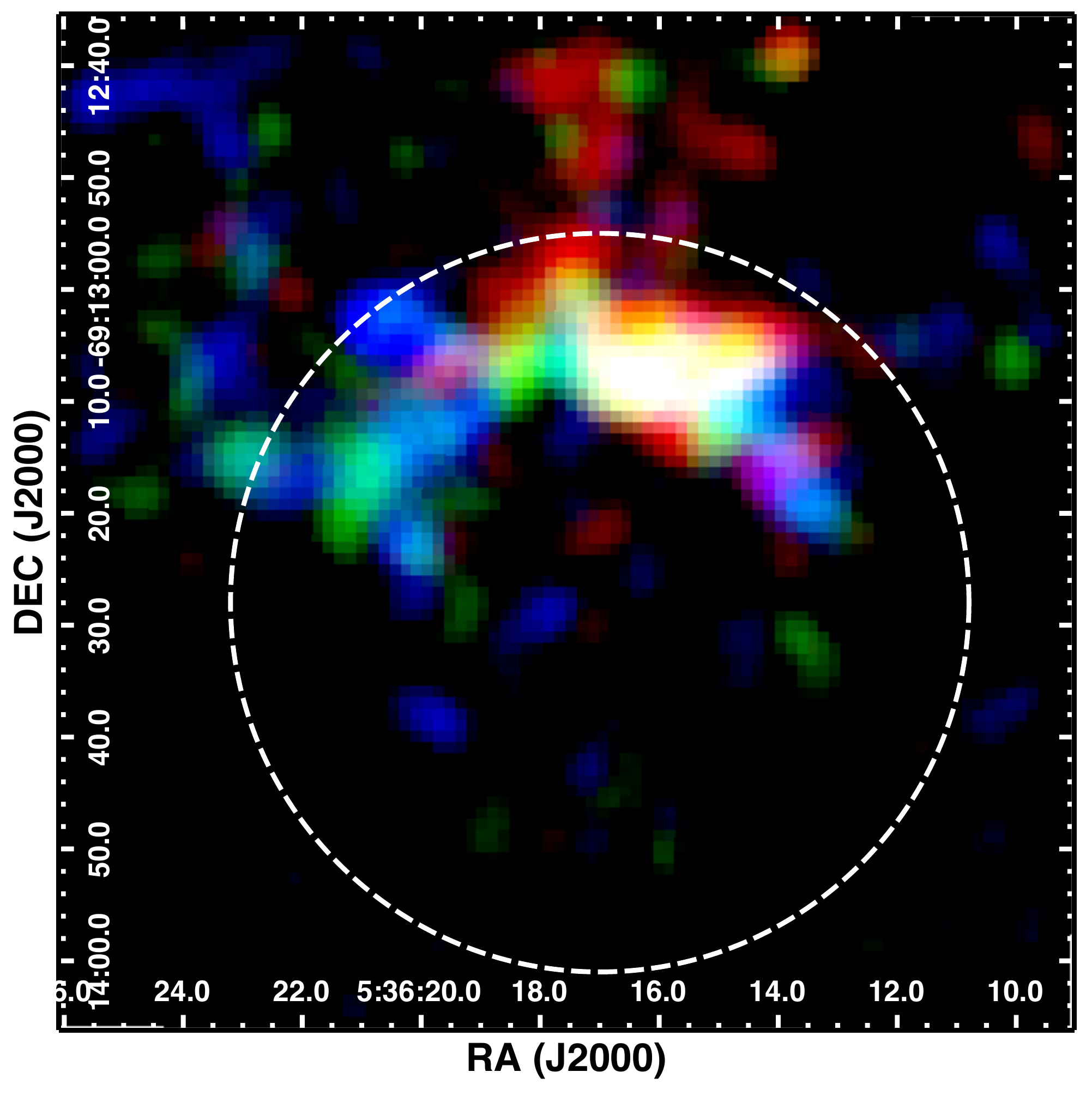}}
\caption{Combined 99+18~ks \chandra\ image of \SNR. Red = $0.5-0.7$~keV (strong O lines), green =$0.7-1.1$~keV (Ne lines), and blue = $1.1-2$~keV (Mg and Si lines). The images are binned by a factor of 2 to improve count statistics and smoothed with a Gaussian kernel of $2\arcsec$. The white dashed circle indicates the dimensions of the remnant determined in Section \ref{x-ray-imaging}. 
}
\label{chandra}
\end{center}
\end{figure}

 To investigate the distribution of the ejecta we made use of the available \chandra\ data (see Section \ref{chan-red}). Since the ACIS-S aimpoint of the observation was SN~1987A, \SNR\ is located $\sim5'$ away on the front-illuminated S4 chip. This results in a degradation of spatial resolution to $\sim2\arcsec$ ($\sim0.5$~pc at the LMC distance). This is still superior to the \xmm\ observations. We created exposure corrected images in the $0.5-0.7$~keV (O lines), $0.7-1.1$~keV (Ne lines), and $1.1-2$~keV (Mg and Si lines), an RGB composition of which is shown in Fig. \ref{chandra}, which were binned by a factor of 2 to improve count statistics and smoothed with a Gaussian kernel of $2\arcsec$. The shell structure detected in the \xmm\ data is also evident here, however, we can already see that the north-western region contains more O and/or swept-up ISM than in the north-east and that our assumption of a representative well-mixed ejecta is an over-simplification. Unfortunately, with only $\sim700$ background subtracted counts for the visible northern arc, a robust spectral analysis is simply not possible.  Only a very deep on-axis \chandra\ observation will allow for a detailed analysis of the distribution of ejecta in \SNR\ and lead to a better estimate of the progenitor mass.

\par For our spectral fits we assumed contributions from swept-up ISM and ejecta emission. We found that \SNR\ is most likely in the ejecta dominated stage of its evolution. Many of the best studied Galactic SNRs are currently in this phase (e.g., SN~1006, Tycho, Kepler, and Cas~A). Neither the free expansion nor Sedov-Taylor solutions are appropriate to describe the evolution of the remnant during the ejecta-dominated phase as both the swept-up mass and ejecta mass must be considered. The analytical solution for the smooth transition from free-expansion to the Sedov phase was given by \citet{Truelove1999}. The characteristic radius, time, and mass of the SNR system are determined using their equations 1, 2, and 3, assuming a uniform ambient ISM ($n=0$ case). To calculate these characteristic values for \SNR\ we needed to determine the initial explosion energy ($E_{0}$), ambient mass density ($\rho_{0}$), and the ejecta mass ($M_{ej}$).~$E_{0}$ was simply taken as the canonical $1\times10^{51}$~erg. Since \SNR\ appears to be immersed in a HII region, we can assume that the ISM is consistent with that of the warm-phase ($T\sim10^{4}$~K) and the ambient number density ($n_{0}$) is $\sim0.1$~cm$^{-3}$. The swept-up mass for $n_{0}\sim0.1$~cm$^{-3}$ and a radius $r=8 (\pm1)$~pc is $7 (\pm1)$~M$_{\sun}$, consistent with the transition phase of \SNR. If the ambient density were an order of magnitude higher (i.e., $n_{0}\sim1~\rm{cm}^{-3}$), the mass swept-up by the remnant is $74 (\pm9)$~M$_{\sun}$, which would dominate the ejecta and the remnant would be well into the Sedov phase. The parameter $n_{0}$ is related to $\rho_{0}$ through $\rho_{0} = n_{0}\mu_{0}$, where $\mu_{0}=1.4m_{p}$ is the mean mass per nucleus. Thus, $\rho_{0} = 2.3\times10^{-25}$~g~cm$^{-3}$. 

\par From the ejecta abundance ratios of our assumed model fits, we determined the likely mass of the stellar progenitor to be either $\sim18$~M$_{\sun}$ or as high as $\gtrsim40$~M$_{\sun}$. At the LMC metallicity, an $\sim18$~M$_{\sun}$ star will spend most of its post-main sequence lifetime in the blue-supergiant (BSG) phase \citep{Schaerer1993}. Assuming that the star sheds its H envelope leading up to the SN event then $M_{ej}\sim7$~M$_{\sun}$ (Yields Table 2013). Using these values we determined the characteristic parameters of \SNR\ and calculated various remnant properties using the shock trajectory parameters of \citet{Truelove1999} for the $n=0$,~$s=10$ case (see their Table 6) where the $s$ value is appropriate for the ejecta distribution of a BSG progenitor. We determined a value of $\sim9$~pc for the radius at which the SNR will transition ($r_{tr}$) from the free-expansion to the Sedov phase. We estimated the radius of the remnant to be $8~(\pm1)$~pc, suggesting that the remnant is on the verge of the transition. The time at which the $r = r_{tr}$ was determined to be $t_{tr}\sim2.2$~kyr, which is a relatively long time to the transition but in keeping with the expansion into a low density ISM. If \SNR\ is on the boundary of the transition, then $t_{tr}$ must approximately represent its age. For a $\sim40$~M$_{\sun}$ star, the same treatment results in $t_{tr}\sim4.9$~kyr and $r_{tr}\sim12$~pc for an ejecta mass of $\sim18$~M$_{\sun}$ (Yields Table 2013), however, this ejecta mass assumes no fall-back which is likely not to be the case. In this situation, the remnant is in the very early stages of the transition and the resulting $t_{tr}\sim4.9$~kyr is an upper limit to the age. Hence, from the progenitor mass estimates, we set a likely age range of $2.2-4.9$~kyr for \SNR. We caution however that this age range determination is subject to our assumptions. In addition the model of \citet{Truelove1999} does not take into account effects such as the Rayleigh-Taylor instability, thermal conduction, magnetic fields, or cosmic ray acceleration, all of which may alter the dynamical evolution of the remnant.

With a likely progenitor mass of $\gtrsim18$~M$_{\sun}$ and emission lines of O, Ne, and Mg in the ejecta component of its spectrum, \SNR\ appears to fall into the ejecta-dominated O-rich SNR class. These remnants are of particular importance as the stellar interior fragments are exposed providing direct probes of stellar evolution and nucleosynthesis models. It has been suggested by, e.g., \citet{Blair2000} that some ejecta-dominated O-rich SNRs result from Type~Ib explosions of massive Wolf-Rayet stars. The LMC contains a small number of such remnants, namely \object{N132D} \citep{Lasker1978} and \object{SNR~0540$-$69.3} \citep{Clark1982}. The SMC also contains a small population, e.g., \object{1E~0102.2$-$7219} (\object{E0102}) \citep{Seward1981} and \object{0103-72.6} \citep{Park2003}. These objects are also noteworthy due to their extraordinarily high X-ray luminosities with N132D and E0102 being the brightest SNRs in the LMC and SMC, respectively. Progenitor mass estimates for these SNRs are $\sim60$~M$_{\odot}$ for N132D \citep{Blair2000} and $\sim30$~M$_{\odot}$ for E0102 \citep{Fink2006}. Morphologically, \SNR\ is most similar to E0102 \citep{Seward1981}. The X-ray morphology of E0102 presents as a ring of ejecta with fainter emission behind the blast-wave due to shock-heated ISM. \SNR\ is similar in this regard assuming our adopted spectral models are representative of the physical composition of the remnant. \SNR\ is larger than E0102 with their radii of $\sim8$~pc and $\sim6$~pc, respectively, of similar age [likely $2.2-4.9$~kyr and $2\pm0.55$~kyr \citep{Fink2006}, respectively], and fainter in X-rays [$\sim10^{35}$~erg~s$^{-1}$ and $\sim10^{37}$~erg~s$^{-1}$ \citep{Gaetz2000}, respectively]. The most likely reason for these differences is the relatively tenuous environment into which \SNR\ appears to be expanding. Therefore, \SNR\ seems to be large analogue of the early evolution of an O-rich SNR.
\par The X-ray shell morphology and spectral signatures derived from our assumed models, suggest a CC SNR is responsible for the extended emission. Multi-wavelength tracers for the SNR are not so clear-cut however. The classic optical signature is the strength of their [\ion{S}{ii}] lines relative to H$\alpha$, with an [\ion{S}{ii}]/H$\alpha$~$> 0.4$ characteristic of SNR emission \citep{Mathewson1973}. \citet{Mathewson1985} found no indication of an [\ion{S}{ii}]/H$\alpha$ ratio consistent with shock ionisation by an SNR from any region of \SB\ meaning there is no optical signature for an SNR at the position of \SNR\ (see also Section~\ref{optical-analysis}) . We repeated this analysis with the MCELS data and the [\ion{S}{ii}]/H$\alpha$ ratio image is shown in Fig.~\ref{ratio}. \citet{Dopita1977} demonstrated that this characteristic ratio was the result of SNR shocks, with velocities 100~km~s$^{-1}$, collisionally ionising and exciting an ambient ISM of sufficiently high density, and subsequent emission line cooling. It is also possible that overrun clouds containing secondary shocks capable of producing a the characteristic [\ion{S}{ii}]/H$\alpha$ signature could exist, even if the primary shock is travelling at $\sim1000$~$km~s^{-1}$. We have shown that \SNR\ is most likely in the transition phase between the free-expansion and Sedov phases and, thus, the SNR shock velocities should be of the order of 1000~km~s$^{-1}$. Also, we inferred that the density of the ambient medium is quite low ($n_{0}\sim0.1$~cm$^{-3}$). Hence, we would not expect to see the characteristic [\ion{S}{ii}]/H$\alpha$ emission from swept-up ISM, but emission due to secondary shocks could be present. However, this would likely be overwhelmed by contamination by the photoionisation emission from the shell of \SB, making any identification difficult. Rather, we expect that the optical emission is dominated by emission lines of the O-rich ejecta (e.g., \ion{O}{ii} and \ion{O}{iii}), such as observed in E0102 \citep{Blair2000}. 
\par We found a relatively steep radio spectral index of $\sim-0.7$ for most of the SNR (see magenta circle in Fig.~\ref{rc-spcmp}), which is consistent with a young remnant \citep[see examples of][]{Bozzetto2014}. However, this value is not only constrained to the immediate vicinity of the remnant so it is unclear as to whether this radio spectral index is representative of \SNR\ or the result of contamination by \SB. In addition, we do not find the typical radial magnetic field (which is a property of younger remnants) around \SNR\ (see Fig.~\ref{rc-pol}), so this may infer possible compression of the region by \SB, rotation through a dense medium, or that this polarisation is simply not from the SNR. For these reasons, we cannot definitively associate any radio emission with \SNR.

\subsection{Non-thermal X-ray emission}
\label{nt-emission}
Our analysis of the deep \xmm\ data has revealed that a non-thermal component is present in all regions of \SB, not just from the bright shell regions. This verifies that the higher emission levels from the shell is a limb-brightening effect. This is further supported by the multi-wavelength morphology (see Section \ref{mwm}) as the hard X-ray shell is highly correlated with the H$\alpha$ and radio shells.


\par YB09 showed that the non-thermal emission mechanism in the shell is most likely synchrotron via their rejection of a simple power law compared to an \srcut\ model for shell~C. Thus, the photon spectrum, and therefore the underlying electron energy spectrum, was observed to roll-off. For their fits, the authors assumed a range of radio spectral indices appropriate for young SNRs. Ideally, we would like to use the radio data in combination with the X-ray data to fit the synchrotron spectrum, however this is problematic for shell~C given the significant thermal contamination at radio wavelengths. Instead, we turn our attention to shell-B which contains the second brightest region of non-thermal X-ray emission in \SB. In addition, shell~B is largely free of thermal radio contamination and we thus assume that the radio emission is entirely synchrotron. We also assume that the hard X-rays are due to synchrotron emission. Hence, we can create a spectral energy distribution (SED) of the photon spectrum due to the underlying relativistic electron population. 

\par
\begin{figure}[!h]
\begin{center}
\resizebox{\hsize}{!}{\includegraphics[trim= 0.05cm 6cm 8cm 0.5cm, clip=true, angle=0]{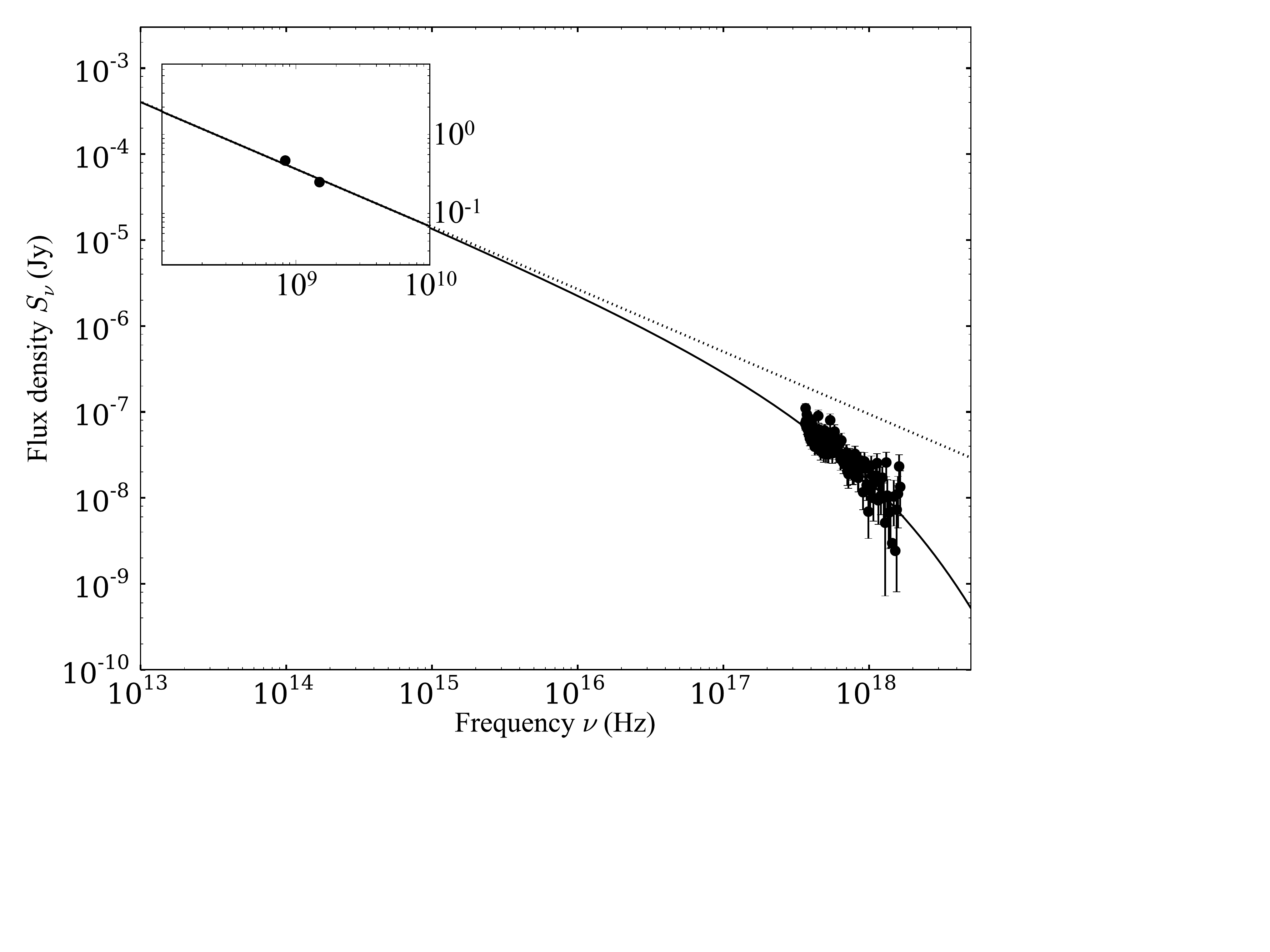}}
\caption{Spectral energy distribution of synchrotron emission from shell~B of \SB. The solid line shows the cut-off power law fit with the dotted line showing the extrapolation of the straight power law. The radio fluxes are at 36~cm and 20~cm, and are shown in the inset to reduce the axis scales of the plot. The X-ray data are the $1.5-7$~keV range.
}
\label{sync}
\end{center}
\end{figure}

 For the radio points, we measured the integrated flux density at 36~cm and 20~cm from shell B. Images at both wavelengths were convolved to the same resolution. This resulted in integrated flux density measurements of 420~mJy at 36~cm and 224~mJy at 20~cm. For the X-ray points, we extracted a spectrum from shell~B from the EPIC-pn data of Obs.~ID~0601200101, the deepest of the EPIC-pn observations. We then extracted and subtracted an adjacent background region to ensure as much as possible that only X-rays due to \SB\ were present. We confined our analysis to the $1.5-7$~keV energy range as below this the thermal emission in shell-B becomes significant. The radio and X-ray data points are shown in Fig.~\ref{sync}.

\par We initially fitted the SED with a straight power law. However it was immediately clear the X-ray fluxes were much lower than would be expected from the radio data. Thus, we introduced a cut-off electron distribution of the form $N_{e}(E) = KE^{-\alpha}e^{-E/E_{max}}$, where $E$ is the electron energy, $E_{max}$ is the cut-off energy, $\alpha$ is the spectral index, and $K$ is a constant. We make the assumption that each electron emits all its energy at its characteristic frequency (the $\delta$-function approximation) and thus the resulting photon spectrum cuts off as $e^{-(\nu/\nu_{max})^{1/2}}$ \citep{Reynolds1998}. This cut-off function provides a much better fit to the SED of shell B with $\alpha = 0.75 (\pm0.02)$, and $\nu_{max}=3.1 (\pm0.7) \times 10^{17}$~Hz. The fit is shown in Fig.~\ref{sync} along with the extrapolation of the straight power law. In this case, $\nu_{max}$ is the characteristic frequency of a photon emitted by an electron at the maximal energy of the electron distribution $E_{max}$. This value is also dependent on the magnetic field and is given by the equation for characteristic frequency $\nu_{c} = 1.82 \times 10^{18} E^{2} B$ \citep{Reynolds1998}. From this equation, we estimated $E_{max}$ in terms of the magnetic field to be $E_{max}~[(B/10~\rm{\mu G})]^{1/2} \sim 80$~TeV. The fact that the spectrum is observed to roll-off is further evidence that the non-thermal X-ray emission from \SB\ is synchrotron in origin.




The main issue with a synchrotron interpretation for the hard shell is the supply of relativistic electrons. In the SNR interpretation of BU04 and YB09, these particles are naturally produced by the strong shocks and associated processes. However, as already discussed, the SNR case should not be possible and there is no multi-wavelength support to this scenario (e.g., [\ion{S}{ii}]/H$\alpha$ ratio, clear SNR radio shell, fast-moving material). SW04 discussed the possibility that the particles may arise from either the wind-termination shock or the bubble turbulence, though suggesting that these either may not be feasible or too inefficient. However, the suggestion of the bubble as the source of the particles is not without merit.

\par \citet[][and references therein]{Parizot2004} showed that the bubble is capable of efficiently supplying a population of non-thermal particles all the way up to $10^{17}$~eV \citep[see also][]{Bykov2001a} through repeated acceleration of low energy particles via turbulence and magnetohydrodynamic (MHD) waves in a superbubble interior caused by the strong stellar wind-wind interactions, interior SNR shocks, and high density clumps of material in the bubble. In addition, \citet{Bykov2001} determined that $20-40$\% of the kinetic energy supplied by the superbubble can be transferred to low energy non-thermal particles and the efficiency is time dependent, peaking after $\sim3$~Myr \citep[see][for an application to the \object{DEM~L192} superbubble]{Butt2008}. Observational evidence for this superbubble particle acceleration process was recently found through $\gamma$-ray emission from the \object{Cygnus superbubble} detected by $Fermi$ \citep{Ackermann2011}.

\par SW04 estimated the integrated stellar wind luminosity to be $(1-7)~\times~10^{37}$~erg~s$^{-1}$ for the 26 O-stars in \SB. In addition, the 7 known WR stars located in \SB\ \citep{Testor1993} are also currently supplying a significant amount of mechanical energy via their fast, dense winds. We adopted mass loss rates and velocities from \citet{Leitherer1997} for the known WR-types and corrected the mass loss rates for the metallicity of the LMC according to \citet{Crowther2007}. This yielded a combined WR luminosity of $\sim5\times10^{38}$~erg~s$^{-1}$. The WR lifetimes are $\sim7\times10^{5}$~yr \citep{Leitherer1997}, so, averaged over the age of the bubble (taken as 4~Myr to be consistent with SW04) they supply $\sim8\times10^{37}$~erg~s$^{-1}$. In addition, SW04 estimated that 5--6 SN have occurred in \SB. Assuming the canonical $10^{51}$~erg input per explosion, this corresponds to an average input of  $(4-5)\times10^{37}$~erg~s$^{-1}$. Thus, the total averaged energy input by the stellar population and SNe is $(1-2)\times10^{38}$~erg~s$^{-1}$. Then, from \citet{Bykov2001} we have $(2-8)\times10^{37}$~erg~s$^{-1}$ transferred to non-thermal particles at peak efficiency. Some fraction of these are electrons which eventually diffuse out to the superbubble shell and if captured in the magnetic field (probably $\sim10~\mu$G assuming a compressed ISM) they can radiate via the synchrotron emission, though it is unclear how efficient a process this would be. The total observed X-ray luminosity of the non-thermal component of the shell is $\sim10^{36}$~erg~s$^{-1}$ which is an order of magnitude less than the energy of the non-thermal particles. Thus, the particle flux from the bubble could potentially explain the observed synchrotron emission.

\subsubsection{Why \SB?}
We have addressed the non-thermal X-ray emission of \SB\ using a multi-wavelength approach and found that a synchrotron origin is most likely. However, one must also consider why \SB\ and no other superbubble in the LMC exhibits such a bright non-thermal shell morphology. If we assume that the non-thermal emission must be due to high energy particles produced in the bubble then shouldn't all superbubbles exhibit similar properties? The answer could simply be that we are observing \SB\ at exactly the right time. \citet{Bykov2001} showed that the efficiency of non-thermal particle production in a superbubble is time dependent, peaking at about 3~Myr, which is near the age of \SB\ (assumed to be 4~Myr). In addition, given that \SB\ is currently at the stage of containing a high mass stellar population, including several WR stars, and interior SNRs, then the energy available for particle production is quite high. For our calculation of the particle production in Section~\ref{nt-emission}, we assumed that the input energy is averaged over the age of the bubble. However, this must be an oversimplification as the onset of SNe must cause spikes in shock energies and turbulence in the interior as the strong shock propagates through the bubble. The current energy input from stellar winds alone in \SB\ is $(5-6) \times 10^{38}$~erg~s$^{-1}$, dominated by the WR population. This is many times the current stellar input of other LMC SBs with values of $\sim1 \times 10^{38}$~erg~s$^{-1}$ for LH9 in \object{N~11} \citep{Maddox2009}, $\sim6 \times 10^{37}$~erg~s$^{-1}$ in \object{N~51D} \citep{Cooper2004}, $\sim7 \times 10^{37}$~erg~s$^{-1}$ for \object{N~70} \citep{Rod2011,DeHorta2014}, $(1-2) \times 10^{38}$~erg~s$^{-1}$ in \object{N~158} \citep{Sasaki2011}, and $\sim9 \times 10^{37}$~erg~s$^{-1}$ in \object{N~206} \citep{Kavanagh2012}. In addition, we have presented evidence of a recent SN near the eastern shell wall in \SB. Although the eastern blast wave of this SNR interacted with the shell wall, the western side propagated into the bubble, adding to the energy available for particle production. For these reasons we suggest that \SB\ is currently undergoing a phase of high energy particle production. If this is the case then other superbubbles must also undergo such stages in their evolution and the energy losses due to the particle production and non-thermal processes must be considered in their overall energy budgets which could alleviate the superbubble growth-rate discrepancy, as suggested by \citet{Butt2008}.

\section{Summary}
\label{sum}
We present an analysis of the large amount of \xmm\ data available for \SB, supplemented by X-ray data from \chandra, optical emission line data from the MCELS, and radio continuum data from ATCA and MOST. The results of our analysis can be summarised as follows:

\begin{enumerate}
\item We detected substantial thermal X-ray emission from the east of \SB. We analyse the superbubble thermal emission, and determine plasma temperatures in the range $kT = (0.17-0.46)$~keV with overabundances O, Ne, and Mg.  Such $\alpha$-enrichment is evidence for a recent CC SNR interaction with the shell. 

\item The new SNR \SNR\ is identified through its clear shell morphology in the $1-2$~keV band. The shell morphology is extraordinarily circular with a north-south brightness gradient. There is no obvious indication for optical or radio emission associated with the SNR. We suggest that \SNR\ is most likely located outside of \SB\ since we would not expect to observe a shell morphology had the blast wave propagated through the SB interior. In addition, the brighter emission from the north of the SNR suggests it is evolving into a higher density medium than in the south, which again is counter-intuitive to a location in the bubble. We determine a radius of $\sim8(\pm1)$~pc. Our X-ray analysis with assumed physical models shows that the remnant is most likely ejecta-dominated with strong lines of O, Ne, Mg, and Si. Based on the derived ejecta abundance ratios, we determine the likely mass of the stellar progenitor to be either $\sim18$~M$_{\sun}$ or as high as $\gtrsim40$~M$_{\sun}$, though the spectral fits are subject to simplifying assumptions (e.g., uniform temperature and well-mixed ejecta). With this progenitor mass range, we set a likely age range of $2.2-4.9$~kyr for \SNR.

\item Using the \xmm\ data we detect non-thermal X-ray emission from all regions of \SB, not just from the bright shell as previously reported, verifying that the higher emission levels from the shell is a limb-brightening effect. This is further supported by the multi-wavelength morphology as the hard X-ray shell is highly correlated with the H$\alpha$ and radio shells. We find that the non-thermal X-ray emission can be fitted equally well with power-law or \srcut\ models. X-ray and radio data are used to produce an SED for the north-eastern shell region of \SB\ which is the second brightest region of non-thermal X-ray emission, but is free of thermal contamination of the radio spectrum. We find that an exponentially cut-off synchrotron model is required to fit the SED with $\alpha = 0.75 (\pm0.02)$, and $\nu_{max}=3.1 (\pm0.7) \times 10^{17}$~Hz. We estimate the maximum energy of the underlying electron distribution in terms of the magnetic field to be $E_{max}~[(B/10~\rm{\mu G})]^{1/2} \sim 80$~TeV. The fact that the spectrum is observed to roll-off is evidence that the non-thermal X-ray emission from \SB\ is synchrotron in origin, which was previously suggested by YU09. However, we argue that this synchrotron emission is not due to an expanding SNR but rather to non-thermal particles produced in the bubble interior being captured in the magnetic field of the shell, which then radiate via the synchrotron emission. We show that the bubble is capable of supplying the required particle flux. We argue that \SB\ is currently undergoing a phase of high energy particle production due to its high-mass stellar population and possibly a recent interior SNR.

\end{enumerate}

\acknowledgements We wish to thank the anonymous referee for the constructive suggestions to improve the paper. This work made use of the XMM-Newton Extended Source Analysis Software. Cerro Tololo Inter-American Observatory (CTIO) is operated by the Association of Universities for Research in Astronomy Inc. (AURA), under a cooperative agreement with the National Science Foundation (NSF) as part of the National Optical Astronomy Observatories (NOAO). We gratefully acknowledge the support of CTIO and all the assistance which has been provided in upgrading the Curtis Schmidt telescope. The MCELS project has been supported in part by NSF grants AST-9540747 and AST-0307613, and through the generous support of the Dean B. McLaughlin Fund at the University of Michigan, a bequest from the family of Dr. Dean B. McLaughlin in memory of his lasting impact on Astronomy. We used the karma software package developed by the ATNF. The Australia Telescope Compact Array is part of the Australia Telescope which is funded by the Commonwealth of Australia for operation as a National Facility managed by CSIRO. P.J.K. acknowledges support from the Bundesministerium f\"{u}r Wirtschaft und Technologie/Deutsches Zentrum f\"{u}r Luft- und Raumfahrt (BMWi/DLR) grants FKZ 50 OR 1209 and FKZ 50 OR 1309, and P.\,M. from grant FKZ 50 OR 1201. M.S. acknowledges support by the Deutsche Forschungsgemeinschaft through the Emmy Noether Research Grant SA2131/1-1. 

\vspace{20mm}

\bibliographystyle{aa}
\bibliography{refs.bib}

\onecolumn
\begin{appendix}
\section{Spectral fit figures}
\label{app-fig}
\vspace{-5mm}

\begin{figure}[h!]
\begin{center}
\resizebox{\hsize}{!}{\includegraphics[trim= 0cm 0cm 0cm 0cm, clip=true, angle=0]{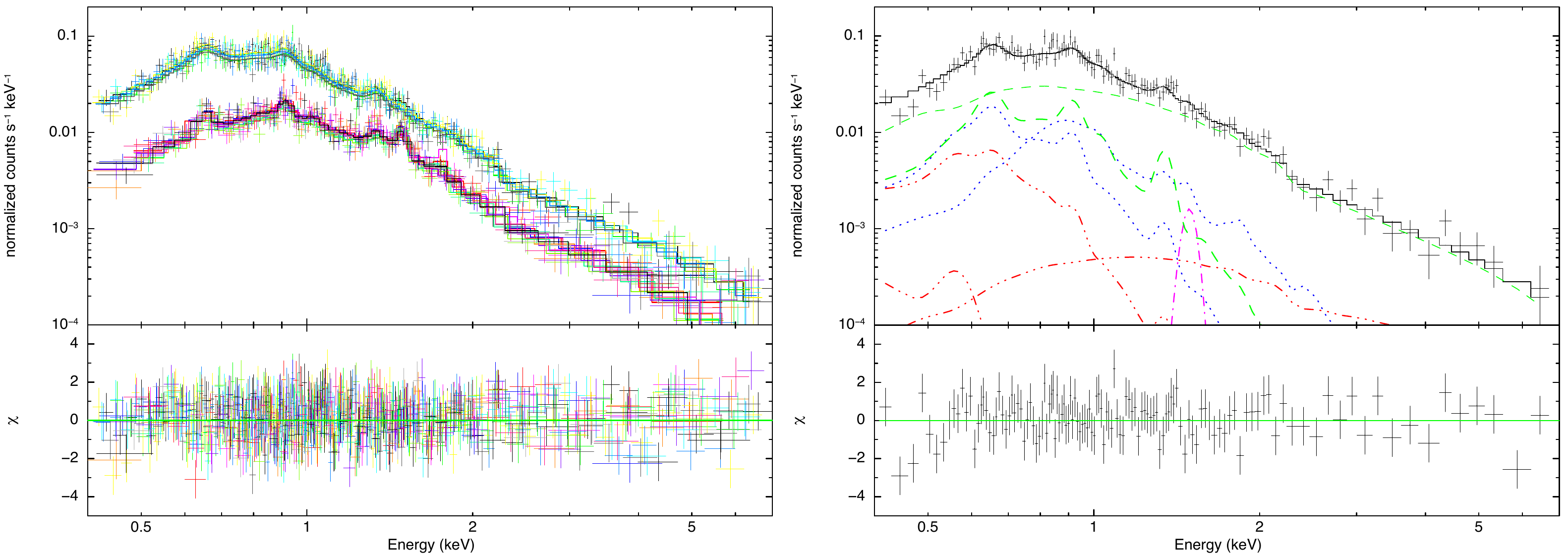}}
\caption{\textit{Left}: Simultaneous spectral fits to the EPIC spectra of B1. \textit{Right}: EPIC-pn spectrum of \SNR\ from Obs.~ID~ 0601200101 (deepest EPIC-pn observation) with additive model components shown. The red dash-dot-dot-dot lines represent the AXB components, the magenta dash-dot line shows the instrumental fluorescence line, blue dotted lines mark the LMC ISM, and green dashed lines represent the source components (source model = \texttt{vapec+pow}). Best-fit parameters are given in Table~\ref{fit-results}. 
}
\label{b1-spectra}
\end{center}
\end{figure}

\begin{figure}[h]
\begin{center}
\resizebox{\hsize}{!}{\includegraphics[trim= 0cm 0cm 0cm 0cm, clip=true, angle=0]{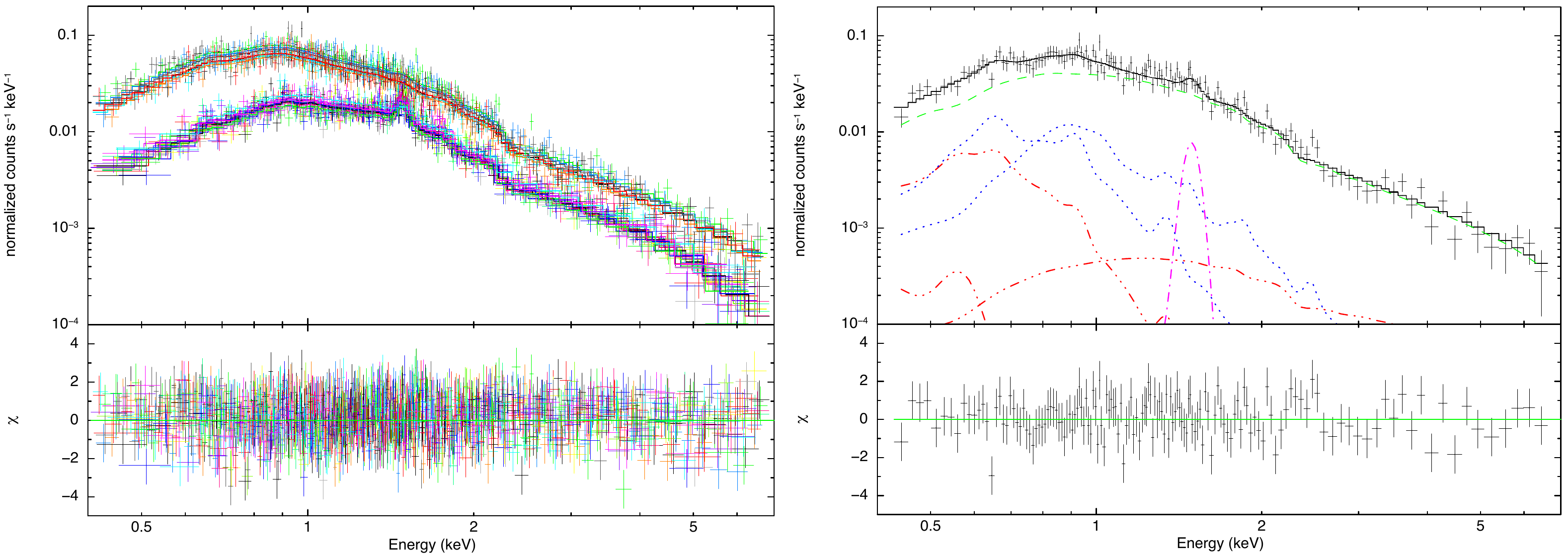}}
\caption{As in Fig.~\ref{b1-spectra} for B2 (source model = \texttt{pow}). Best-fit parameters are given in Table~\ref{fit-results}.
}
\label{b2-spectra}
\end{center}
\end{figure}

\begin{figure}[!h]
\begin{center}
\resizebox{\hsize}{!}{\includegraphics[trim= 0cm 0cm 0cm 0cm, clip=true, angle=0]{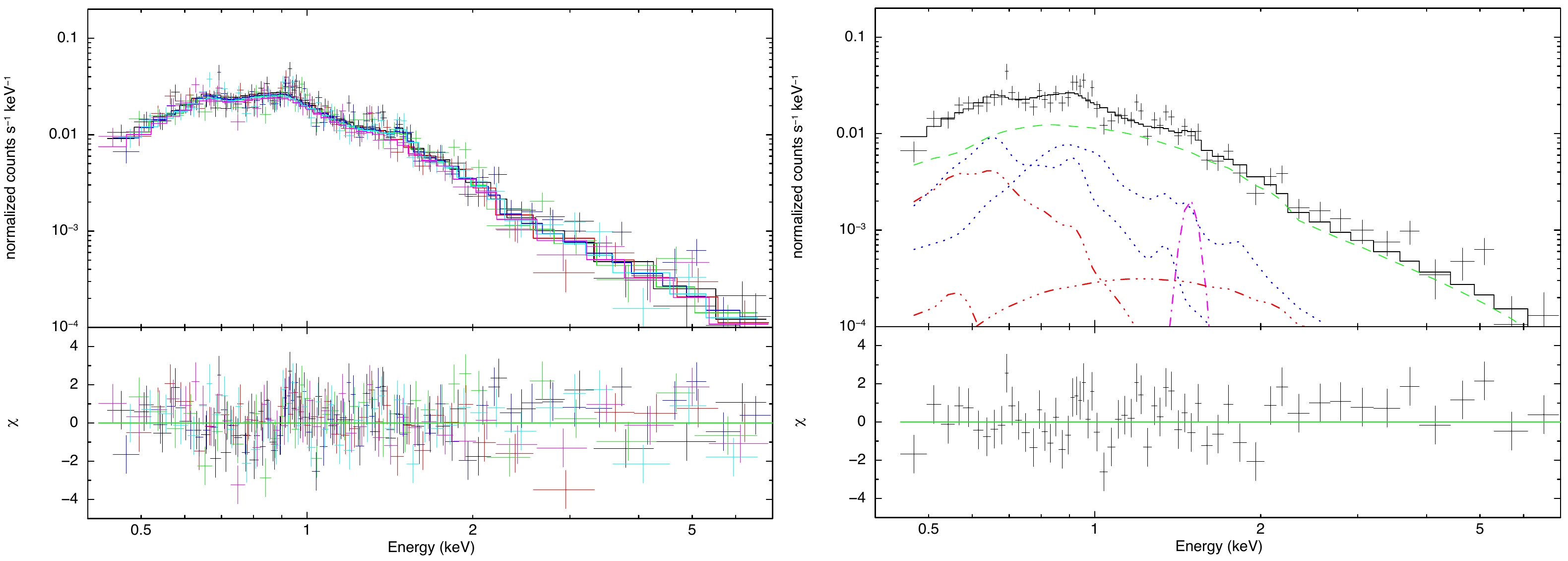}}
\caption{As in Fig.~\ref{b1-spectra} for B3 (source model = \texttt{pow}). Best-fit parameters are given in Table~\ref{fit-results}.
}
\label{b3-spectra}
\end{center}
\end{figure}

\begin{figure}[!h]
\begin{center}
\resizebox{\hsize}{!}{\includegraphics[trim= 0cm 0cm 0cm 0cm, clip=true, angle=0]{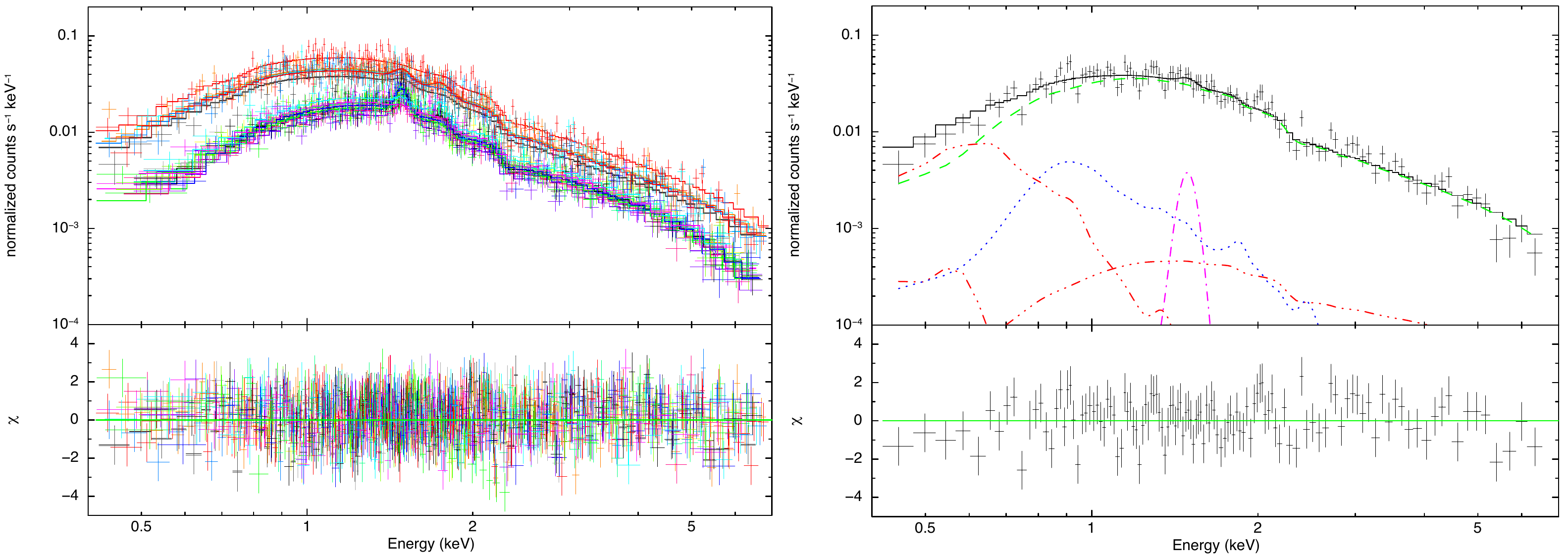}}
\caption{As in Fig.~\ref{b1-spectra} for C1 (source model = \texttt{pow}). Best-fit parameters are given in Table~\ref{fit-results}.
}
\label{c1-spectra}
\end{center}
\end{figure}

\vspace{-10mm}

\begin{figure}[!h]
\begin{center}
\resizebox{\hsize}{!}{\includegraphics[trim= 0cm 0cm 0cm 0cm, clip=true, angle=0]{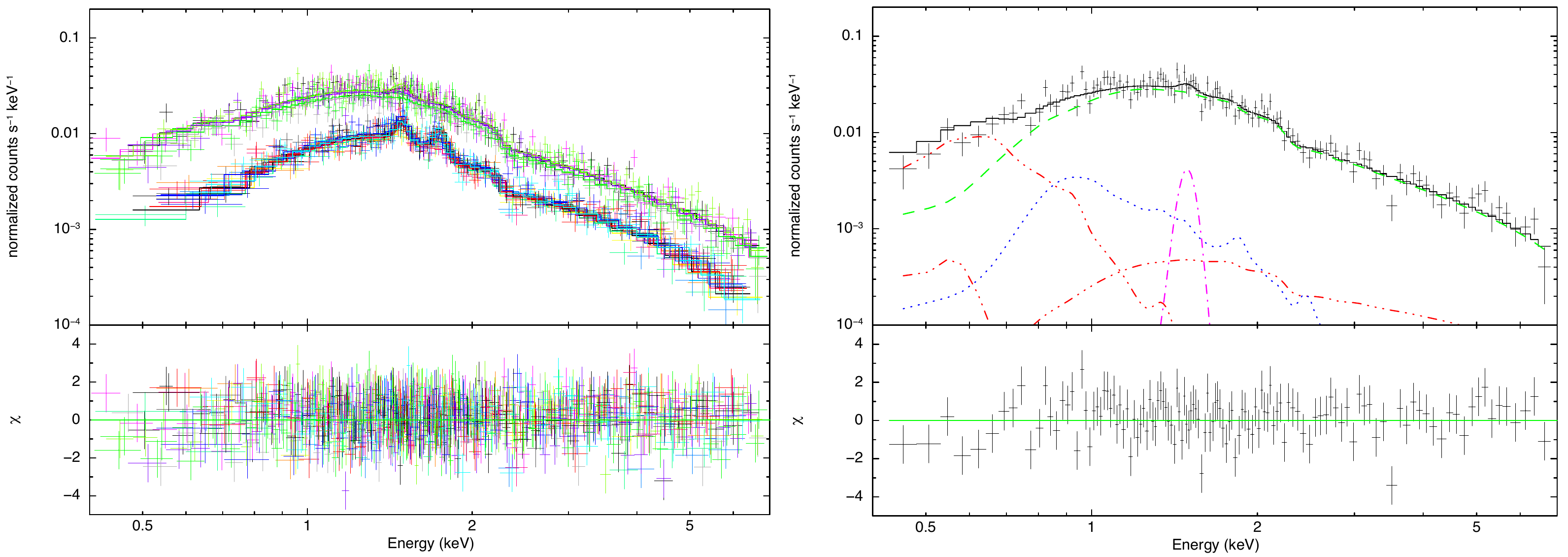}}
\caption{As in Fig.~\ref{b1-spectra} for C2 (source model = \texttt{pow}). Best-fit parameters are given in Table~\ref{fit-results}.
}
\label{c2-spectra}
\end{center}
\end{figure}

\vspace{-10mm}

\begin{figure}[h]
\begin{center}
\resizebox{\hsize}{!}{\includegraphics[trim= 0cm 0cm 0cm 0cm, clip=true, angle=0]{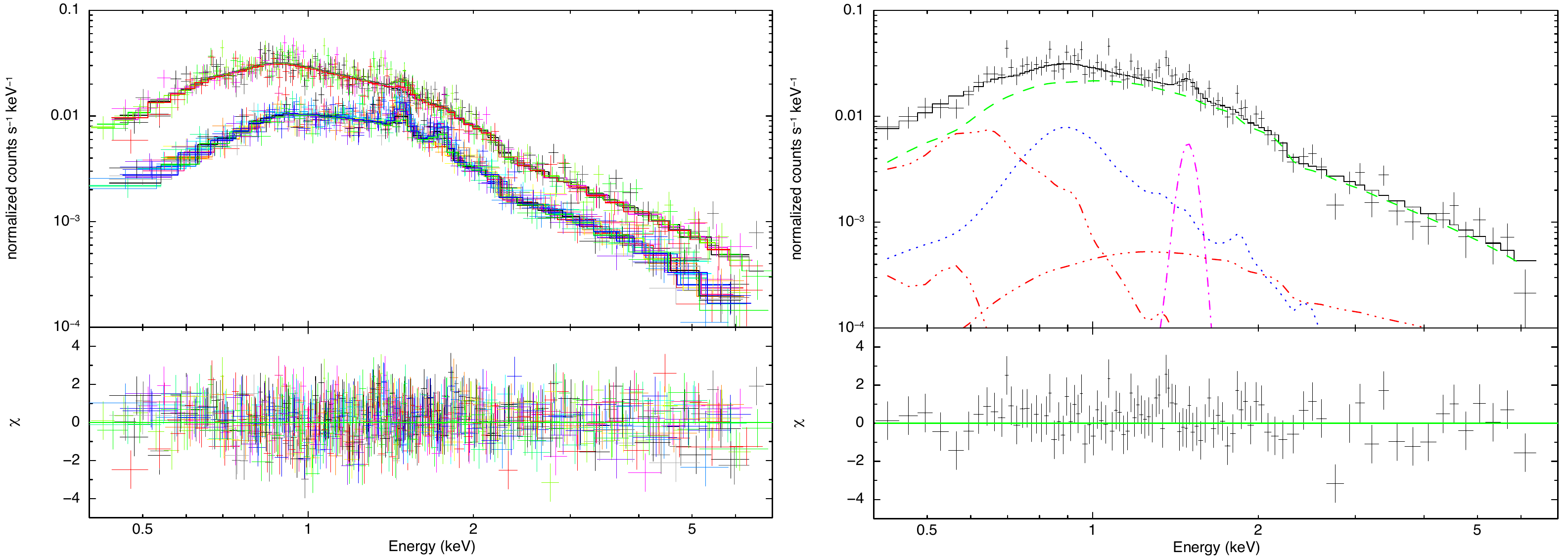}}
\caption{As in Fig.~\ref{b1-spectra} for D (source model = \texttt{pow}). Best-fit parameters are given in Table~\ref{fit-results}.
}
\label{d-spectra}
\end{center}
\end{figure}

\vspace{-10mm}

\begin{figure}[!h]
\begin{center}
\resizebox{\hsize}{!}{\includegraphics[trim= 0cm 0cm 0cm 0cm, clip=true, angle=0]{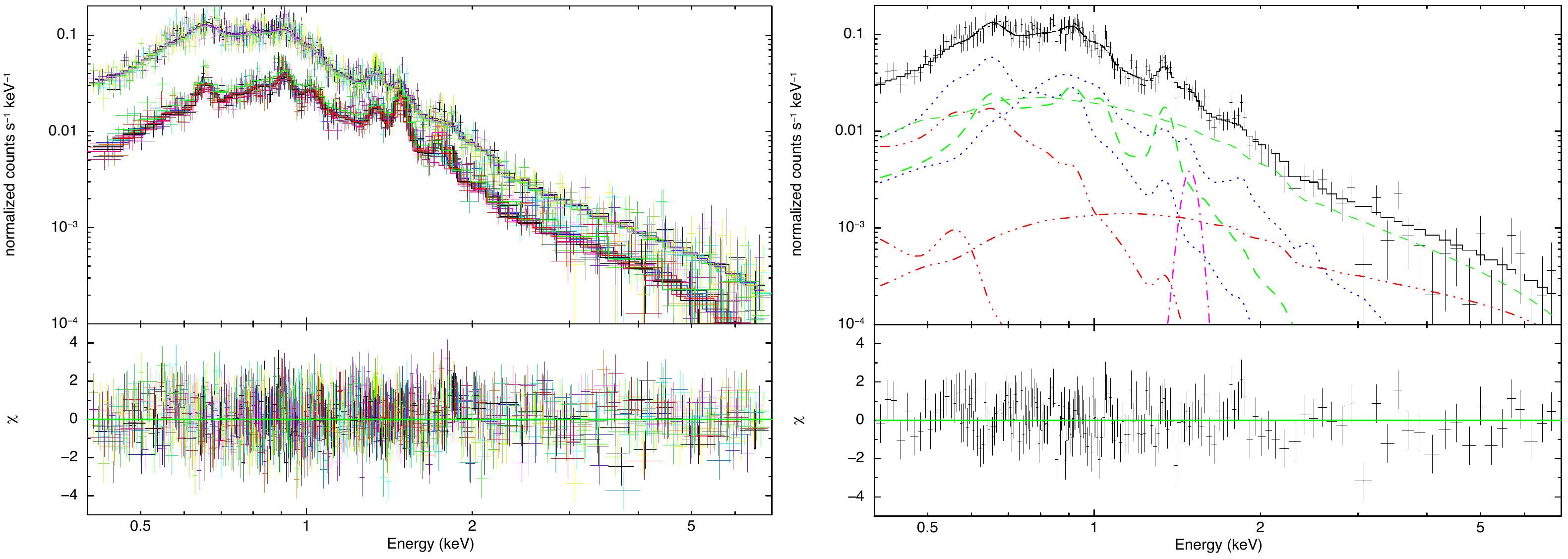}}
\caption{As in Fig.~\ref{b1-spectra} for I1 (source model = \texttt{vapec+pow}). Best-fit parameters are given in Table~\ref{fit-results}.
}
\label{i1-spectra}
\end{center}
\end{figure}

\vspace{-10mm}

\begin{figure}[h]
\begin{center}
\resizebox{\hsize}{!}{\includegraphics[trim= 0cm 0cm 0cm 0cm, clip=true, angle=0]{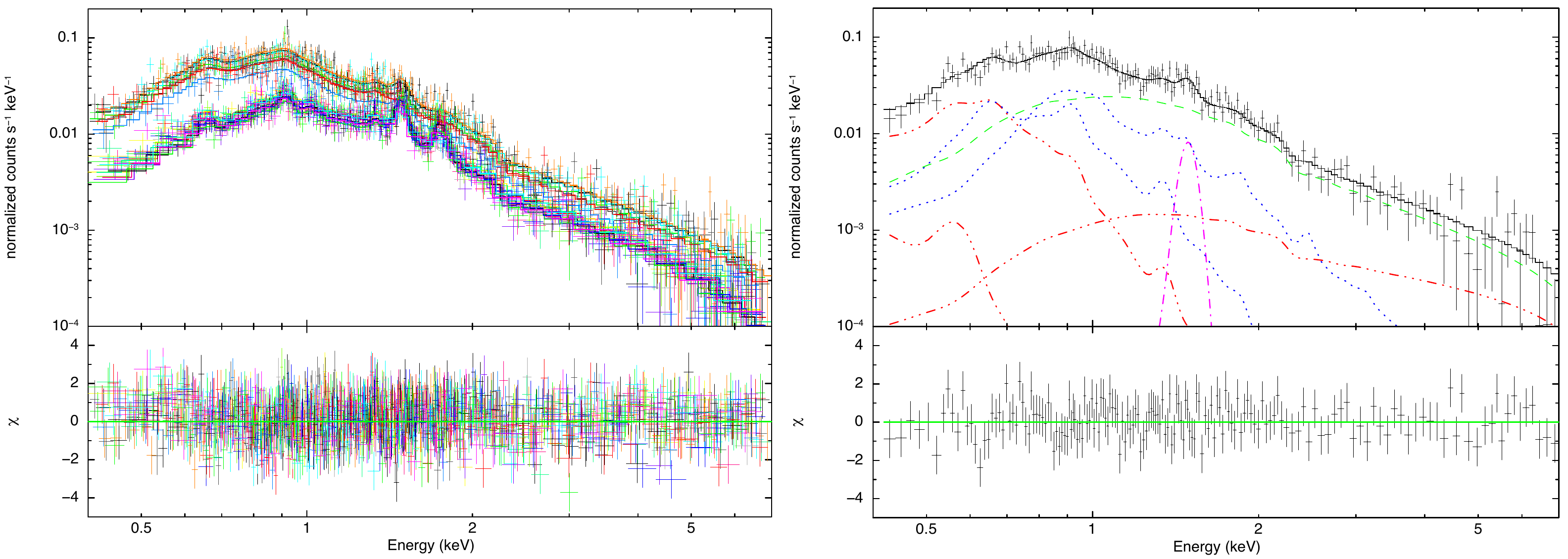}}
\caption{As in Fig.~\ref{b1-spectra} for I2 (source model = \texttt{pow}). Best-fit parameters are given in Table~\ref{fit-results}.
}
\label{i2-spectra}
\end{center}
\end{figure}

\vspace{-10mm}

\begin{figure}[h]
\begin{center}
\resizebox{\hsize}{!}{\includegraphics[trim= 0cm 0cm 0cm 0cm, clip=true, angle=0]{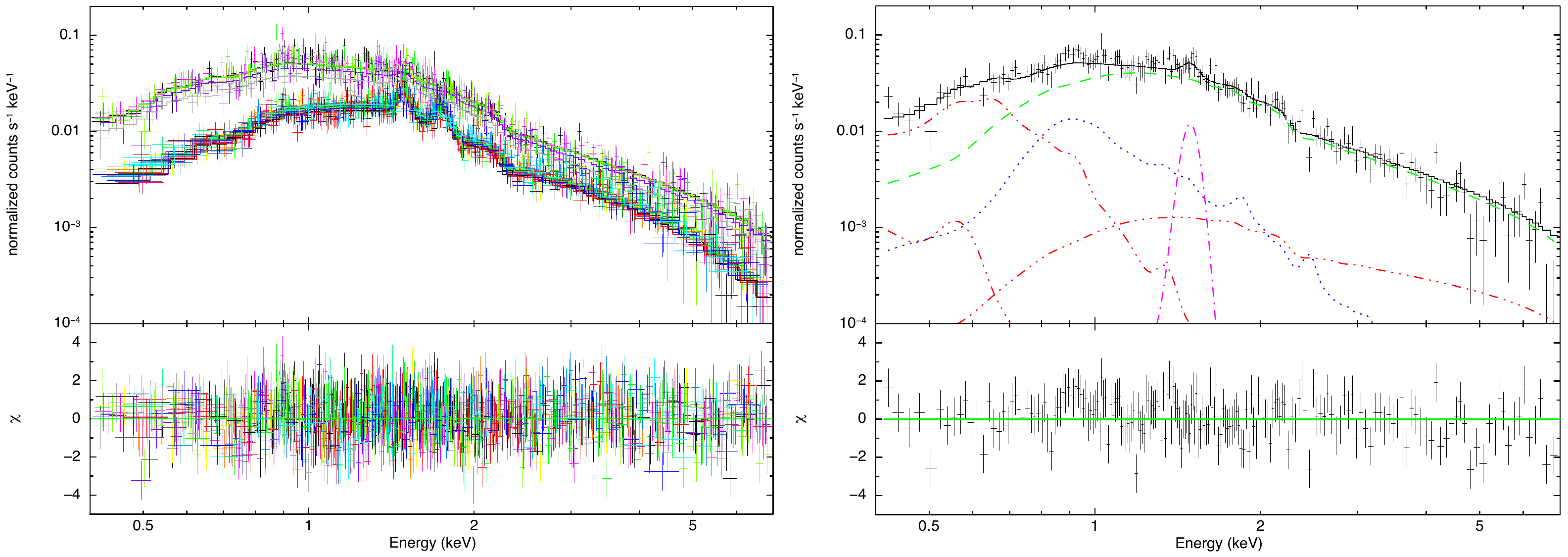}}
\caption{As in Fig.~\ref{b1-spectra} for I3 (source model = \texttt{pow}). Best-fit parameters are given in Table~\ref{fit-results}.
}
\label{i3-spectra}
\end{center}
\end{figure}
\end{appendix}

\end{document}